\documentclass{aa}  

\usepackage{graphicx}
\usepackage[export]{adjustbox}

\usepackage{txfonts}
\usepackage{color}
\def\etal{{et al.\ }}

\def\kms{\ {\mathrm{km\,s^{-1}}}}

\def\nveltot{19,708} 
\def\nvel{18,146} 

\def\ngaltot{11,910} 
\def\ngal{10,719} 

\def\newvel{5,084} 
\def\newvelgal{4,617} 
\def\newgal{2,585} 

\def\SSGtot{63,471} 
\def\SSGfaint{294} 

\begin{document} 

%
\title{A redshift database towards the Shapley Supercluster region\thanks{Based on observations made at Las Campanas Observatory (Chile) and Cerro-Tololo Interamerican Observatory (Chile).}}

\subtitle{}

\author{Hern\'an Quintana\inst{1}
   		\and 
        Dominique Proust\inst{2}
        \and 
        Rolando D\"unner\inst{1}
        \and 
		Eleazar R. Carrasco\inst{3}
        \and 
        Andreas Reisenegger\inst{1}
        }

\institute{
   		Instituto de Astrof{\'\i}sica, Facultad de F{\'\i}sica, Pontificia Universidad Cat\'olica de Chile, Casilla 306, Santiago~22, Chile. Mail: hquintana@astro.puc.cl
        \and
        GEPI, Observatoire de Paris, F92195 MEUDON PRINCIPAL CEDEX, France.
        \and
        Gemini Observatory, NSF’s National Optical-Infrared Astronomy Research Laboratory, Casilla 603, La Serena 1700000, Chile.
        }


    \date{February 12, 2020}

 
\abstract
{We present a database and velocity catalogue towards the region of the Shapley Supercluster based on \nvel~measured velocities for \ngal~galaxies in the approximately 300 square degree area between 12h43m00s $<$ R.A. $<$ 14h17m00s and $-23^\circ 30'00'' > $ Dec $ > -38^\circ 30'00''$. The data catalogue contains velocities from the literature found until 2015. It also includes \newvel~velocities, corresponding to \newvelgal~galaxies, observed by us at Las Campanas and CTIO observatories and not reported individually until now. Of these, \newgal~correspond to galaxies with no other previously published velocity measurement before 2015. Every galaxy in the velocity database has been identified with a galaxy extracted from the SuperCOSMOS photometric catalogues. We also provide a combined average velocity catalogue for all \ngal~galaxies with measured velocities, adopting the SuperCOSMOS positions as a homogeneous base. A general magnitude cut-off at R2=18.0 mag was adopted (with exceptions only for some of the new reported velocities). In general terms, we confirm the overall structure of the Shapley Supercluster as found in earlier papers. However, the more extensive velocity data show finer structure, to be discussed in a future publication. 
}

\maketitle
\titlerunning{Shapley Supercluster velocities}
%

\section{Introduction}

Superclusters of galaxies are the largest structures that can be identified in large redshift surveys of galaxies or in catalogs of clusters of galaxies, of sizes $\sim 100\,\mathrm{Mpc}$. With densities a few times larger than the average density of the Universe, they are already in the non-linear regime, but still far from being virialized (contrary to clusters of galaxies). Thus, their complex spatial and velocity structure still largely reflect their initial conditions, and their limits are ill-defined and dependent on arbitrary criteria such as density thresholds or linking lengths. 

D\"unner \etal (2006) proposed a physical definition of superclusters as the largest structures that will remain bound in spite of the accelerated expansion of the Universe, and thus will in the future collapse to form stable, virialized, spherical clusters that separate from each other at exponentially increasing velocities (e.~g., Araya-Melo \etal 2009). In a spherical model, the outermost shell of a bound structure of mass $M$ approaches an asymptotic radius $(3GM/\Lambda)^{1/3}$ (where $G$ is Newton's gravitational constant and $\Lambda$ is Einstein's cosmological constant), corresponding to an average enclosed matter density twice that associated with the cosmological constant, $\rho_\Lambda\equiv\Lambda/(8\pi G)$. In the present Universe (with density parameters $\Omega_m=0.3$ and $\Omega_\Lambda=0.7$), the enclosed matter density is still 1.69 times higher than this asymptotic value, corresponding to 2.36 times the current critical density or 7.87 times the current average matter density of the Universe (Chiueh \& He 2002; D\"unner \etal 2006). The density threshold set by this criterion is rather high compared to those usually chosen to define superclusters, implying that most of the mass of observationally defined superclusters is actually not bound, and the physical definition yields smaller superclusters, as found, e.~g., by Chon \etal (2013). In order to distinguish these two definitions, Chon \etal (2015) proposed to call the latter ``\emph{superstes} clusters'' (``survivor clusters''), as they will survive the accelerated expansion.

In the nearby Universe, at redshifts $z<0.1$, several superclusters have been identified, including our own. Tully \etal (2014) have proposed an extension of our own supercluster, where the Local Group and Milky Way reside, to be  greatly increased due to the discovery of new galaxies in the traditional Zone of Avoidance. These numerous galaxies form a connection between the Virgo supercluster and other nearby structures, including 13 Abell clusters, forming a larger entity named the Laniakea supercluster, with a diameter of 160~Mpc (equivalently, 12,000$\kms$ in velocity) and a total mass of $10^{17}\,M_\sun$. However, the density enhancement and dynamical state of the Laniakea supercluster lead to the prediction (Chon \etal 2015) that the whole structure is unbound and will disperse due to the effect of dark energy in the long-term future (with some substructures or clusters remaining bound).

The more distant \emph{Shapley Concentration} or \emph{Shapley Supercluster} (SSC) at $z\approx 0.05$ was long ago recognized as one of the largest structures in the nearby Universe (Shapley 1930). Melnick \& Quintana (1981) were the first to identify the central cluster, later named A3558, as a bright X-ray cluster. Subsequently, Melnick \& Moles (1987) studied the several major clusters and main groups in the central regions, using data later published by Quintana \etal (1995), where the general supercluster structure was further defined. Its unique position within the $z<0.1$ Universe was pointed out by Raychaudhury (1989) and Scaramella \etal (1989). 

Catalogs of superclusters (Einasto \etal 2003a, 2003b) find the SSC to be the largest and richest supercluster at $z<0.1$ (although Chon \etal 2013 find it to split up into several ``superstes clusters''). As such, it can play an important role in constraining models of structure formation (Sheth \& Diaferio 2011). It has also been invoked as being responsible for the motion of the Local Group with respect to the cosmic microwave background (e.~g., Melnick \& Moles 1987; Kocevski \etal 2004; Courtois \etal 2017). In addition, it is an interesting laboratory to understand the interactions between clusters of galaxies, as well as the evolution of galaxies in high-density regions, but before falling into clusters (Merluzzi \etal 2013, 2016). For such applications, it is necessary to accurately characterize the structure and dynamics of the supercluster, in order to measure its total mass and density distribution and assess the evolutionary state of its different regions. This requires large redshift surveys, which in fact have been carried out over the years (Quintana \etal 2000; Proust \etal 2006; Haines \etal 2018; and references therein).

Over the last few years, the number of velocity measurements in this direction has been rapidly increasing, mainly thanks to multi-object spectroscopic surveys such as FLASH (Kaldare \etal 2003), 6dF (Heath Jones \etal 2009) and many observations by our group. Recently, Haines et al. (2018) presented a deep, highly complete redshift survey of the SSC's core region (21 square degrees or $[17\,\mathrm{Mpc}]^2$) and a detailed analysis of the structure in the core region.

In this paper, we present a compilation of 
velocities in a wide area ($20^\circ\times 15^\circ$), including our previously unreported data. The latter were obtained mainly from a survey carried out over several years with the fiber and multislit spectrographs at the 100'' du Pont telescope at Las Campanas Observatory (LCO), as well as the Hydra spectrograph at the Blanco 4.0\,m telescope at Cerro-Tololo Interamerican Observatory (CTIO). 
Here, we present all the data resulting from these observations, which represent a set of \newvel~new velocities among which \newgal~are newly observed galaxies. Combined with already published redshift sets from several surveys and papers, compiled until 2015, we built up the most extensive velocity database for the SSC area, containing \nvel~velocity measurements for \ngal~galaxies. 

The observing runs and instrumentation used in the spectroscopic observations are described in Sect. 2. In Sect. 3 we present the extraction and build-up of a photometric galaxy catalogue from the SuperCOSMOS Sky Surveys (hereafter SSS)\footnote {\url{http://www-wfau.roe.ac.uk/sss/index.html}.} 
for the approximately 300 square degree area of the sky.
The extended velocity database is described in Sect. 4, while Sect. 5 describes the average velocity and photometry catalogue, obtained from cross-combining the velocities from the database and the photometry and astrometry from the galaxy catalogue derived from the SSS.
This section includes comparisons between the galaxy velocities in common among different data sets, as well as the velocity zero-point shifts required to equalize the combined average catalogue.
In Sect. 6, we discuss the completeness of this average velocity catalogue and analyze the galaxy number density over the whole and the intercluster survey regions. Finally, Sect. 7 describes the observed galaxy density, velocity distribution and general structure of the supercluster.

\section{Observations and instrumentation} \label{Observations and instrumentation}

The selected SuperCOSMOS region extends in Right Ascension (RA) from 12:44:50.557 to 14:17:43.609 and in Declination (Dec) from -38:30:03.41 to -23:27:30.24, containing 60710 objects identified by us as galaxies. The galaxy positions, photometric magnitudes (R1, R2, Bj and I), and morphological properties were inherited from the SuperCOSMOS values.

The spectroscopic observations were carried out using initially the fiber spectrograph and then the WFCCD, both mounted on the 2.5 m du Pont telescope at Las Campanas Observatory, Chile (LCO)
\footnote{Las Campanas Observatory (LCO) is an astronomical observatory owned and operated by the Carnegie Institution for Science (CIS).}, and the HYDRA fiber spectrograph on the 4.0\,m Blanco telescope at Cerro-Tololo Inter-American Observatory, Chile (CTIO)
\footnote{Cerro Tololo Inter-American Observatory is operated by the Associations of Universities for Research in Astronomy under contract with the National Science Foundation.}.

The multi-fiber system first used at LCO consists of a plug plate at the focal plane, with 128 fibers running to a spectrograph coupled to the 2D-Frutti detector, covering a field of view of 1.5 $\times$ 1.5 degrees on the sky. Instrumental details and observing procedures were the same as on earlier reported observations described in Shectman (1989), Quintana \etal (2000), and Proust \etal (2006). 

The WFCCD is a multi-slit drilled bronze mask with useful 22~arcmin $\times$ 22~arcmin field of view, taking two fields per cluster to cover central regions.
Blue grism 400~lines/mm was used for spectra exposure, with a 2k $\times$ 2k CCD, binned 1$\times$1 with gain~1.
We took three or four 900s exposures per field, and He-Ar comparison lamps were taken before and immediately after each set of exposures. 

HYDRA is a multi-object spectrograph with 138~fibers of 2~arcsec diameter on the sky operating in a 40~arcmin diameter field of view. 
It uses a 400~mm camera and SITe 4k $\times$ 2k detector, binning 1$\times$2 with gain~2, noise 3$e^{-}$ with filter GG385. 
The KPGL2 grating resolution was 0.7\AA/pix with a spectral range 3600-7400\AA/mm and tilt $5.74^\circ$. 
We made 3~exposures of 480s, 720s, or 1200s depending on the average brightness of the galaxies in the field. 

Table~\ref{newobs} gives a schedule of the observing sessions and a code for the corresponding observation sets used in the following Database. In Fig. 1 we plot the fields covered by the spectroscopic observations, as explained in its caption.

\begin{table*}[htb]
\caption{Observing sessions and instrumentation used.}
\small
\begin{tabular}{ccccccc}
\hline
\hline
 Observatory & Telescope &   Instrument     &  Spectral   & Dispersion  &   Date     &  Code   \\
             &           &    Detector      &  range      &             &            &         \\
\hline
LCO          &   2.5m    & Fibers           &  3500-7000 \AA  &  1\AA/pix   &  01/1992   &  QC01   \\
LCO          &    ``     &      ``          &      ``     &     ``      &  05/1993   &  QC02   \\
LCO          &    ``     &      ``          &      ``     &     ``      &  03/1994   &  QC03   \\
LCO          &    ``     &      ``          &      ``     &     ``      &  05/1994   &  QC04   \\
LCO          &    ``     &      ``          &      ``     &     ``      &  02/1995   &  QC05   \\
LCO          &    ``     &      ``          &      ``     &     ``      &  03/1995   &  QC06   \\
LCO          &    ``     &      ``          &      ``     &     ``      &  01/1996   &  QC07   \\
LCO          &    ``     &      ``          &      ``     &     ``      &  03/1997   &  LC97   \\
LCO          &    ``     &      ``          &      ``     &     ``      &  03/1998   &   ``    \\
LCO          &    ``     &      ``          &      ``     &     ``      &  05/1998   &   ``    \\
LCO          &    ``     &      ``          &      ``     &     ``      &  03/1999   &   ``    \\
CTIO         &   4.0m    &  HYDRA           &  3196-7980 \AA & 0.7\AA/pix  &  02/2006   &  HYDRA  \\
LCO          &   2.5m    &  WFCCD           &  3800-7600 \AA & 3.0\AA/pix  &  02/2007   &  WFCCD  \\
LCO          &    ``     &      ``          &      ``     &     ``      &  05/2008   &   ``    \\
LCO          &    ``     &      ``          &      ``     &     ``      &  03/2009   &   ``    \\
\hline
\end{tabular}
\label{newobs}
\end{table*}

\begin{figure*}
\begin{centering}
\includegraphics[trim=10 20 0 50, clip, width=2.0\columnwidth]{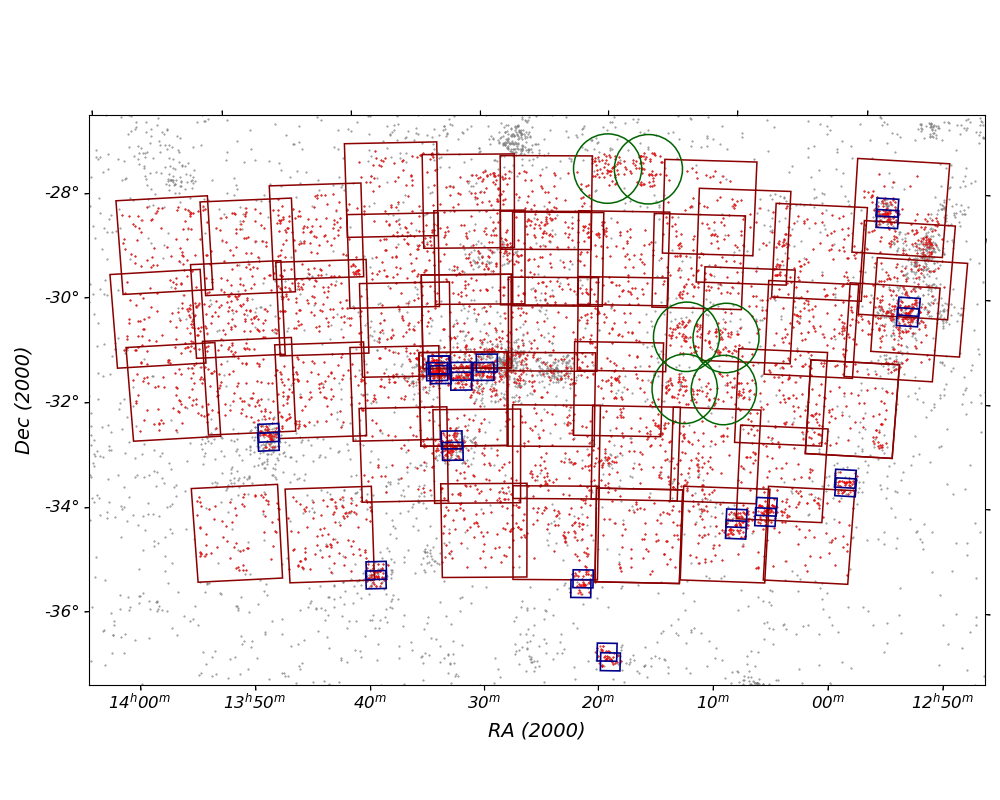}
\par\end{centering}
\caption{Spectroscopic survey fields reported in this paper. Large squares are LCO fiber spectrograph observations. Circles are Hydra fields observed at CTIO and small blue squares are WFCCD fields observed at LCO. The area of this plot is smaller than the total catalog area, chosen to zoom in the spectroscopic surveyed fields. The gray dots represent the positions of galaxies in the velocity catalog.}
\label{survey_areas}
\end{figure*}


From the above observing runs, we obtained a set of \newvel~new velocities for \newvelgal~galaxies (some of them were observed two or more times) in the general direction of the SSC. 

\section{Photometric and astrometric catalogue}

One underlying problem with data collected from many sources over a 
period of many years comes from the (usually small) differences in the positions assigned to the galaxies in different surveys. 
Assuming there are no ambiguities in identification, the different positions quoted in the literature (particularly in older data) make it hard in some cases to cross-identify the velocities in an automatic, homogeneous way. There are also some ambiguities in identification, some due to misprints in listed positions, juxtaposition of (brighter) large galaxies to faint nearby ones, or due to crowding and superposition of galaxies near cluster or groups centers. Older data come from many sources, with varying positional errors, including some of our earlier work (Quintana \etal 1997 and references therein, and in Table 2). If we include HIPASS observations, as we did, 
position errors can reach up to $6'$ (Doyle \etal 2005). 
A valuable tool became available in the South with the access to the SuperCOSMOS Sky Surveys (SSS) facility with homogeneous astrometry and photographic photometry. SSS provides digitized scans of the B(J), R(R1 and R2) and I color band Sky Surveys plates, with a pixel size of 0.7 arcsec. The SSS is available from the Web, including downloads of images and extraction catalogue of objects according to different parameter choices.

The extraction of object catalogues from the SSS was done with the parameter set left at their suggested default values, except image quality, as discussed in what follows.
Two parameters of particular relevance were the Classification flag (1=galaxy, 2=star, 3=unclassifiable, 4=noise), and the Quality flag, specifying an object's image quality (see SSS). Two files were produced, one for CLASS 1 objects, in principle classified as galaxies, and one for CLASS 2 objects, in principle stars. Below we discuss these classifications in more detail. CLASS 3 or 4 objects were extremely few; they were analysed and eventually considered not relevant for the study.

Images and object catalogues were extracted in an area centered near the position of the central cluster A3558: RA(2000) 13:30:00, DEC (2000) -31:30:00. We chose the output with the suggested most useful subset of master IAM (Image Analysis Mode) data as suitable for our purpose. Images were extracted with $30'\times 30'$ size. The object catalogues produce positions, proper motions (disregarded), magnitudes in all four bands, image ellipticity fits, position angle, ellipse area, image quality, and class. The ESO SkyCat software package (\url{http://www.eso.org/sci/observing/tools/skycat.html}) permits to display the survey images overlapped by the object catalogues, each object represented by an ellipse with its center coordinates, area, ellipticity, and position angle parameters. Different colors or symbols also allow simultaneous overlays of several catalogues. Due to the not uncommon presence of blended galaxy images in clusters, we kept both the parent object defined by the common low isophotal contour and the daughter objects, represented by ellipses around the individual galaxies, at higher isophote contours. In the visual analysis, we chose the best isophote to represent each galaxy image, deleting others. We used the standard pairing distance of 3". From many trials by visual inspection, we found that the
suggested default quality flag extraction value 128 missed too many real galaxy images in the neighborhood of bright stars, because of contamination by reflections and spikes. 
Therefore, the quality parameter was chosen as 4096 to extract basically all the galaxy images, from an upper magnitude limit of R2=10 down to a lower limit of R2=18.0 mag, as the red R magnitude is more representative of cluster galaxies. 
However, we found the B images were better for the classification eye checks of galaxy/star/spurious object thanks to the better image quality and contrast of galaxy disks and halos in this band.

The catalogue area extracted extends over 300 square degrees (exactly 301.3), between  12h43m00s  $<$ R.A.(2000)  $<$ 14h17m00s and $-23^\circ 31'00'' > $ Dec(2000) $ > -38^\circ 30'00''$. In this way we covered nearly the complete area of possible SSC candidate clusters (including the WFCCD observations). Several extracted partial catalogues were merged, carefully avoiding duplication of objects on the overlapping border areas. In total, we extracted 1,300 images of $30'\times 30'$ in size (in a $42\times $31 grid, with a 1 arcmin overlap to avoid missing sky areas or objects near the edges) and all 1,300 fields were searched by eye to clean the catalogues by removing spurious objects and repeated isophotal contours.

The $30'\times 30'$ survey image fields were first looked at on the screen in sections of $10'\times 10'$, when the first clean-up of the galaxy (CLASS 1) object catalogue  performed, eliminating obvious reflections and star spikes, plate defects and other noise. For some rather bright galaxies, several elliptical isophotal contours were given, of which we chose the lower, wider area ellipse representative object, that had a center coincident with the galaxy optical center.  Examples of not so faint double stars, identified by a common ellipse as CLASS 1, were deleted. In case of doubts, usually due to faint double star images identified as a single object, we looked at higher magnification. The observed inhomogeneous plate image quality and increasing star numbers and image confusion at lower Galactic latitudes (towards the south), both combined to make the selection easier in the cleaner Northern sections of the area and increasingly harder in the Southern region. The number of spurious objects extracted from the catalogue classified as CLASS 1 was large, in some survey plates several times larger than the actual galaxy number. Needless to say, this was the effect of using the value 4096 as image quality, but also of the increasing number of stars and worse seeing  at lower Galactic latitudes. Thus, at lower Galactic latitudes, where the star density more than doubles that present at higher latitudes, worse seeing in an original plate would produce more objects spuriously classified as galaxies due to image diffusion and blended images of close faint optical star pairs.
These effects introduce an unavoidable systematic effect in the catalog, increasing the uncertainty of its completeness near the limiting magnitude in those fields. In summary, among the objects extracted as CLASS 1 (in principle classified as galaxies) with our chosen parameters, in each $30'\times 30'$ survey image field between 40\% to 85\% of objects (depending on its declination and presence of bright stars) were found not to be galaxies, but different kinds of other objects, as described above. The actual numbers vary widely if the field fell on a cluster or a fairly empty region.

There are additional factors that increase photometric and positional uncertainties of some catalogue objects. These effects are caused by the superpositions of galaxy/star or galaxy/galaxy images, not properly deblended by the catalogue software, 
and depend on the density of galaxies and stars, which vary with field positions (if they fall over clusters) and with Galactic latitude. Therefore, across the whole supercluster area surveyed this effect is quite variable. Of order 15\% of 
objects catalogued as galaxies have photometric or positional parameters distorted (in most cases only slightly) by superposition of two galaxy or galaxy/star images.

The procedures applied were as follows: 
\begin{itemize}
    \item[a)] Galaxy/star super or juxtapositions. When the dominant part of the image was a star (producing most of the magnitude), over a very faint galaxy, these catalogue objects were eliminated. When the star was a perturbation of similar or smaller magnitude to the galaxy image, these objects were retained in the galaxy catalogue (with the resulting parameter values). Naturally, the photometry values would be in error, but a significant galaxy could be retained. However, the fitted ellipse was also distorted and its center position altered. No sharp limit between these cases could be used, as it depended on relative magnitudes and center distances, galaxy shape, and seeing. In a few cases, when the center of the galaxy isophote was off by several seconds and there was a velocity measured, we corrected the center position to the center of the galaxy to aid the correct identification.
    \smallskip
    \item[b)] Galaxy/galaxy superpositions. When the de-blending procedure worked, we kept both separate ellipse objects as galaxies. However, in many cases this did not happen. When one galaxy was much fainter than the companion, we kept the catalogue object as given. If we did not have velocities for either component, we did similarly. For galaxies brighter than around R2=17.0, if two  galaxy images in contact were represented by one entry and we had some component velocity, we extracted a catalogue over a small area trying different extraction parameters, to check if merged or parent/daughter images were available that solved the problem. However, when this procedure did not work, we added another entry by hand, adjusting both positional and angle parameters to represent the two galaxies as separate objects (duplicating the photometry). These procedures were important in the centers of clusters, when bright dumbbell galaxies or several superimposed galaxies were present (most with velocities measured). In this way, we tried to keep the galaxy identification correct and complete (to correlate with the velocities).
\end{itemize}

A rare, but non-negligible situation arose occasionally, first discovered when 
the position assigned to a velocity measurement fell on top of a relatively bright elliptical galaxy image not appearing in the CLASS 1 objects. We looked at the CLASS 2 catalogue and found a corresponding object. Because we found several other cases, with or without velocities, we looked on the screen at the whole catalogue of CLASS 2 objects overlapped on the survey images. Occasionally, we found obvious (fairly bright) galaxies wrongly classified as stars, of order 2-3 per square degree. These were fairly bright E0 ellipticals, with clear diffuse edges. We transferred these entries to the galaxy catalogue but kept their CLASS 2 for easy later reference. As a further step in the verification of the galaxy identifications we simultaneously plotted the NED database\footnote{The NASA/IPAC Extragalactic Database is operated by the Jet Propulsion Laboratory, California Institute of Technology, under contract with the National Aeronautics and Space Administration.} with different symbols over the survey images. This was of particular help near the cut-off photometric limit R2=18.0 mag.

In total, we extracted more than 1,184,000 objects of all classes. 
The number of CLASS 2 (star) entries was roughly one million. The number of original CLASS 1 (galaxy) objects was close to 173,000.  
All of them were checked individually by eye, as a majority were false or problem classifications (as a result of using the quality image parameter 4096, as discussed). 
The number of galaxy objects kept was slightly more than 60,400.
Thus, the number of real galaxies was around 30\% of the CLASS 1 entries. 
To these entries, we added a few hundred additional galaxy entries.
A few of these were generated ``by hand'' in the case of double or superimposed galaxies, as explained, or star images that turned out to be galaxies and moved to this catalogue. A few additional objects were galaxies fainter than R2=18.0 that had velocities measured by us (with a value within or close to the Shapley velocity range). For the statistical analysis in Sec. 6, these faint galaxies were ignored by using the magnitude cut-off at R2=18.0 mag. 
 
A further point needs to be discussed. The SSS software was designed to identify fairly faint galaxies. When galaxies are extended over several arcmin (i.e., galaxies with velocities less than $\sim$ 3,000$\kms$, not members of the SSC), and particularly of low surface brightness, with several bright spots, the software produces isophotes centered at those bright spots, with no isophote describing the whole galaxy, neither well centered. Since these galaxies have photometry in the literature, we chose the partial isophote that appeared best centered on the galaxy nuclei, to help the identification algorithm. For these extended area galaxies, the SSS optical parameters are not a good description of the image (see discussion in the HIPASS optical identification catalogue, Doyle et al. 2005).


The total number of galaxy entries in the derived SuperCOSMOS-based catalogue is \SSGtot\ (with \SSGfaint~fainter than R2=18.0 mag with new velocities). Hereafter, we call this the ``Shapley area SSS galaxy catalogue'', or SASSSG catalogue, for short. A large majority of its galaxy entries have visually correct parameters (centers, magnitudes, ellipticities, and position angles).

\section{Velocity database}

By combining our new velocity measurements with existing ones from the literature, we produced a large velocity database containing \nveltot~velocity measurements corresponding to \ngaltot~galaxies, over an area slighly bigger than the one finally chosen as defined earlier, and covering most of the SSC. When limited to the 300 square degrees, we obtained the quoted \nvel~velocities and \ngal~galaxies finally retained in the database.

Our search of the literature is not intended to be complete, particularly for velocities $<$ 6,000$\kms$. 
We added source references as we found them for one or another reason. 
We realize that several literature papers include previous data along with their own measurements, and thus the same velocity data point could inadvertently be considered several times, coming from different sources. 
A difficulty comes from the fact that in the literature the individual source of a velocity measurement is not always given. When the duplication is clear (because it coincides with an older value already included in the database), we eliminated the entry, keeping the original measurement. A comment was added in these entries. Besides, there are references to databases dynamically varying in time, such as ZCAT (\url{https://heasarc.gsfc.nasa.gov/W3Browse/all/zcat.html}) or NED. The FLASH survey uses these references as only source or to average values with their own measurements (\url{http://ned.ipac.caltech.edu/}). If there has been an update of the those databases, the old values source may be lost or are hard to find.
Moreover, When an author quotes a velocity value which is an average of his measurement and an older NED and/or ZCAT value, it is hard to know which both those velocities were (unless NED or ZCAT has kept the old value). More complex is the case when the NED or ZCAT later updates their values exactly to the average given by the same reference. 
Here we give individual velocity values with references for each of them, trying, as far as possible, to quote the original measurement only. 
However, when a literature average is given, we listed it as well, since it can contain new partial information from velocities unavailable today, for the reasons just given (and we warn about this in the comments). 
Generally we have also tried to give in the database the original numbers for the several parameters of the literature entry, as far as given by the authors.  

We used the reported coordinates of each velocity measurement to compare with the SASSSG catalogue, grouping together multiple measurements of the same object, while identifying its visual counterpart.
This was an iterative process done by finding the closest SASSSG galaxy counterpart to each entry in our preliminary database, up to a maximum allowable distance of 20'', and comparing cross-match velocities for consistency.
The SSS images where used for visual inspection in case of conflicts encountered by our automatic analysis software.

After a first matching run, we identified all the SASSSG objects that were assigned to more than a single counterpart in the velocity catalog, producing a list of duplicate matches. 
To resolve these cases, we began by accepting the closest match within each set to be correct, and then asked whether the other velocity measurements corresponded or not to that same SASSSG counterpart.
If no other SASSSG object was found within 20'' of the other velocity measurement, and if the velocities lied within 500$\kms$, then we merged the two velocity measurements as different measurements of the same object.
If otherwise there was another SASSSG candidate closer than 20'', or if velocities differed by more than 500$\kms$, then we marked those for visual inspection.
This was done overlapping survey images and positions on the screen, both from the SASSSG and velocity object positions.
When two or more velocity positions fell close to a visual galaxy, with greatly discordant velocities, we took several steps. 
First, we did a search of the original references, to discard typos, precession errors or relevant updates, and obvious identification errors. 
Some were resolved at this level. 
Then, we evaluated the given redshifts against the apparent galaxy image brightness, compactness and morphological details, to try to discard the less sensible velocity. 
Simultaneously, we considered the quoted velocity errors or spectra quality factors (if given) and similar discordant problems in the same reference, to try to decide which velocity to keep for the database. 
In nearly all cases, we could allocate a preferred velocity. 
For completeness of the raw data, velocities not considered for further calculation were kept in the velocity database, but flagged accordingly. 
In a few rare cases, a velocity position fell midway between two galaxy images with similar magnitude and morphology. 
Similar steps were followed, looking in the references and finding charts or descriptions. 
In most cases the problem was solved. 
Only in two or three cases, the uncertainty remained and these data were deleted from the database. 

One persistent problem was posed by contact and/or close double galaxies. First, the COSMOS software rarely produced two separate catalogued objects when the distance between centers was less than around 10"-20". 
As described earlier, we introduced a second image, adjusting the centers and ellipticities correspondingly. 
Moreover, the velocity data from different authors usually listed a velocity for one component only, with position errors of the same magnitude as the distance between centers. 
If the velocities of the components were not very different, it was difficult to allocate the correct identification. 
We did our best using all the information available (if at least one author gave the two velocities, or we tried to choose the brightest as the observed one, or trust the accuracy of the relative positions of one author, etc.). 
For some of these cases, we had to manually force the identifications in the automatic identification procedure. 
No more than two or three cases of allocation remain uncertain. 

Unidentified velocities were mostly associated to blank spots in the SSS images, due to position errors (as was the case with some IRAS objects (Allen \etal 1991, or HIPASS galaxies) or misprints in the galaxy positions in the published velocity tables that made them fall on blank sky or near a star. 
In most of the latter cases, we could find the real target galaxy and corrected the misprint. 
Manually resolved conflicts were incorporated into the database manually in the next matching run. 
Most often, we moved the velocity entry with good positional consistency to the first line of the group of velocities corresponding to one galaxy, which was used for the positional correlation.

The reduced database contained a grand total of \nvel~velocities corresponding to \ngal~galaxies.

Our velocity database groups different measurements for each galaxy, cross-identified as described above. For traceability, we kept the original position entries (unless a misprint was detected or we were forced to use NED positions for identification purposes), followed by all the velocity parameters as published in the corresponding reference or in our own work.
Each entry line gives an individual velocity measurement, including a reference code to identify the reference source and a FLAG (1 or 0) to indicate if the velocity should be used, or not, to compute the average velocity of the object.
Different galaxies are separated by a line containing only an index number.
A sample of these data appears in Table \ref{catalog_sample}, where we also give the list of all velocity references. The comments column contains further information regarding the references, problems with the identification and/or indicate the criteria used or choices made. Discrepant velocity values were also noted. In Appendix I we discuss some individual galaxies that have wrong or complex optical identifications.

\begin{table*}[htb]
\label{catalog_sample}
\caption{Shapley Supercluster velocity Database. A running number separates different galaxies, with one or more velocity entries from indicated reference.
(1) object number or Galaxy ID, with a letter labelling each entry;
(2) survey, field or cluster name;
(3) internal number in original reference or catalog, if available. Otherwise is 1;
(4) Right Ascension (J2000), as given in reference (unless noted otherwise in comments); 
(5) declination (J2000), as given in reference (unless noted otherwise in comments); 
(6) magnitude in original reference (several photometric systems)
(7) heliocentric velocity in $\kms$, in reference; 
(8) associated error in units of $\kms$, in reference; 
(9) R number from Tonry \& Davis (1979) for velocity determination; 
(10) number of emission/absorptions lines (\#l) used to calculate the velocity;
(11) flag (1 if velocity is considered for further analysis or 0 otherwise);
(12) Code for original reference or catalog.
(13) additional comment about the source (see also original reference).
}

\tiny
\begin{tabular}{lllrrrrrrrrll}
\hline
{\bf(1)}     &    {\bf(2)} &  {\bf(3)}   & {\bf(4)} & {\bf(5)} & {\bf(6)} & {\bf(7)} & {\bf(8)} & {\bf(9)} & {\bf(10)} & {\bf(11)} & {\bf(12)} & {\bf(13)}\\
\hline
1 &&&&&&&&&&&& \\
 G124213.2-275210\_a & FLASH         &                 1                           &   12:42:13.220 & -27:52:09.40 &  16.50  &   2187    &   54  &   0.0 & 0.0  &  1 & Ka03  &     F        \\                                         
 G124213.2-275210\_b & 6dFGS         &                 127630    &                     12:42:13.260 & -27:52:10.90 &  16.88 &    2179     &  45  &   0.0 & 0.0 &   1 &  6dFGS    &            \\                                        
2 &&&&&&&&&&&& \\
 G124314.9-310413\_a & 6dFGS         &                 60699     &                     12:43:14.980 & -31:04:12.90 &  16.30 &   15696     &  31  &   0.0 & 0.0  &  1 & 6dFGS &     Q4 sp; NED ref    \\                                
 G124314.9-310413\_b & FLASH         &                 1         &                     12:43:14.890 & -31:04:12.80 &  16.20  &  16599      & 44  &   0.0 & 0.0  &  0 & Ka03 &      F \\                                                
 G124314.9-310413\_c & C12403        &                 5     &                         12:43:11.000 & -31:03:25.00 &   0.00 &   15589    &  120 &    0.0 & 0.0   & 1 & Al91 &      IRAS 12405-3047 \\
3 &&&&&&&&&&&& \\
 G124313.7-343419\_a & FLASH &                         1     &                         12:43:13.760 & -34:34:18.50 &  16.30  &   6026    &  108  &   0.0 & 0.0  &  1 & Ka03    &   F \\
 G124313.7-343419\_b & 6dFGS     &                     126674    &                     12:43:13.800 & -34:34:20.10 &  16.66 &    6047    &   45 &    0.0 & 0.0 &   1 & 6dFGS & \\
4 &&&&&&&&&&&& \\
 G124351.2-321937\_a & FLASH     &                     1         &                     12:43:51.200 & -32:19:37.10 &  15.90 &    9308    &   35  &   0.0 & 0.0  &  1 & Ka03 &      F     \\
 G124351.2-321937\_b & 6dFGS     &                     119214    &                     12:43:51.240 & -32:19:37.30  &  16.21 &    9037  &     45 &    0.0 &  0.0  &   1 & 6dFGS  & \\         
\hline

\end{tabular}
\end{table*}

\section{Combined velocity-photometry catalogue}

The large velocity database permits the extraction of a velocity catalogue, with one average velocity for each galaxy observed in the region.
We obtained this in two steps: 
1) try to correct for the small, in principle systematic, differences in the literature, produced by different instrumentation or measurement procedures. 
2) Do a weighted average of the available velocities for each object. 
The result is a velocity catalogue based on the database. 
From the database, each future worker can calculate its own corrections for a velocity catalogue. 
It is debatable whether the best catalogue should be built taking the best velocity measurement for each galaxy (as the NED or ZCAT were built), or whether a weighted mean is more suitable, as this value would be influenced by poor data, even if properly weighted down. 
However, even extremely accurate data, such as HI velocity measurements, can be uncertain, because the HI velocity might be different from that of the galactic nucleus. Moreover, there is the problem of the correct identification for HI scanning surveys, such as HIPASS (in Appendix I we further discuss the latter problem).

\subsection{Zero-point correction}
In order to combine our new data with already published velocities, we shift all data to a common zero-point. 
This helps to control systematic errors produced by possible zero-point mismatch of radial velocities used by different authors working on different instruments.
As the starting reference set for the zero-point we used the velocity measurements performed by us with the fiber spectrograph at the du Pont telescope with the same detector, including previous ones described in earlier papers (Quintana \etal 1995, observing sessions Q01, Q02, Q03, Q04, Q05, Q06, Q07, M01, M03, and Quintana \etal 2000, observing sessions listed in Table~\ref{newobs}: QC01, QC02, QC03, QC04, QC05, QC06, QC07, LC97, not individually reported before). 
We earlier showed (Quintana \etal 1995 and Quintana \etal 2000) that there were no systematic velocity differences between observations carried out with the fiber spectrograph and the earlier Reticon spectrograph (both built by S. Shectman, Shectman 1989), mounted at the du Pont Telescope, as both used the same Carnegie Image Tube as final detector, used for the much earlier observations with the same telescope reported in (Quintana \etal 1995). 
In most sessions we also carried out repeated observations of some galaxies (as listed in the database), to check that we had a consistent velocity system and to reduce statistical errors.

For each of the other surveys (both our own with the WFCCD and Hydra, and all those from the literature), we identified the galaxies in common with the reference set. 
Starting by the survey with the largest number of velocities in common with the reference set, $N_{com}$, we used the average of the velocity differences between the survey and the reference set, $\overline{\Delta v}$, as our zero-point correction. After correcting the survey by its zero-point, we included it in our reference set, increasing the number of measurements available for comparison with the remaining surveys, and continued the process with the next one with largest $N_{com}$.

Assuming Gaussian measurement errors, the significance of the zero-point correction can be quantified by computing the standard deviation of the velocity differences, $\sigma_{\Delta v}$, such that the error on the mean becomes $\sigma_{\overline{\Delta v}} = \sigma_{\Delta v}/\sqrt{N_{com}}$.
The significance can then be written as $\mathrm{S/N}\equiv |\overline{\Delta v}|/\sigma_{\overline{\Delta v}}$.
Table~\ref{zeropoint} summarizes these results, where $N_{ref}$ is the number of galaxies of the corresponding reference given in column 2. 

\begin{table*}[htb]
\caption{ Zero-point corrections for each of the surveys included in our database with respect to our reference set. The variables are discussed in the text.}

\small
\begin{tabular}{llrrrrrr}
\hline
\hline
 Code    &      References             & $N_{vel}$   & $N_{com}$ & $\overline{\Delta v}$ & $\sigma_{\overline{\Delta v}}$ & S/N &  $\sigma_{\Delta v}$    \\
         &                             &      &      &$\kms$&$\kms$& & $\kms$  \\
\hline
   6dFGS & Heath Jones et al. (2009) and previous releases & 5956 & 1287 &  20 &   3 &  7.1 & 103 \\ 
    Ka03 &       Kaldare et al. (2003) & 1454 & 1140 &   6 &   3 &  1.8 & 118 \\ 
    Sm04 &         Smith et al. (2004) &  678 &  331 &  15 &   4 &  3.4 &  80 \\ 
    Dr91 &             Dressler (1991) &  358 &  295 &  17 &   5 &  3.1 &  91 \\ 
    DW04 &    Drinkwater et al. (2004) &  328 &  240 &  -2 &   7 & -0.3 & 106 \\ 
   Ktg98 &       Katgert et al. (1998) &  303 &  228 &   2 &   5 &  0.4 &  75 \\ 
    DW99 &    Drinkwater et al. (1999) &  295 &  207 &  -2 &   8 & -0.3 & 108 \\ 
    St96 &                Stein (1996) &  385 &  217 & -13 &   5 & -2.5 &  75 \\ 
    Da98 &      da Costa et al. (1998) &  232 &  201 &   4 &   6 &  0.6 &  89 \\ 
    Ba01 &      Bardelli et al. (2001) &  569 &  202 &  17 &   6 &  2.6 &  91 \\ 
    Wi99 &       Willmer et al. (1999) &  156 &  141 &  40 &   6 &  7.0 &  68 \\ 
    QM97 &      Quintana et al. (1997) &  291 &  130 & -14 &  10 & -1.4 & 114 \\ 
    Ba97 &      Bardelli et al. (1997) &  418 &  131 &   1 &  13 &  0.1 & 149 \\ 
    Ba94 &      Bardelli et al. (1994) &  309 &  129 &   6 &  10 &  0.7 & 111 \\ 
    Sm07 &          Smith et al.(2007) &  226 &  139 &  23 &   6 &  4.2 &  65 \\ 
   WFCCD & This work (Las Campanas WFCCD) &  555 &  126 & -35 &  11 & -3.1 & 128 \\ 
    Da86 &      Da Costa et al. (1986) &  109 &  106 &  38 &   7 &  5.2 &  76 \\ 
   Cav09 &          Cava et al. (2009) &  192 &  100 &   7 &   7 &  0.9 &  73 \\ 
    Da87 &      Da Costa et al. (1987) &  102 &   97 &  45 &   7 &  6.3 &  70 \\ 
    Ba98 &      Bardelli et al. (1998) &  174 &   94 &  -5 &  11 & -0.4 & 103 \\ 
    We03 &        Wegner et al. (2003) &   94 &   94 &  36 &   7 &  4.9 &  70 \\ 
   The98 &      Theureau et al. (1998) &   88 &   82 &   9 &   7 &  1.3 &  61 \\ 
    Te90 &        Teague et al. (1990) &   87 &   76 & -33 &  10 & -3.3 &  87 \\ 
    Al91 &         Allen et al. (1991) &   86 &   80 & -48 &  11 & -4.3 & 100 \\ 
    Dr88 & Dressler \& Shectman (1988) &  101 &   81 &  66 &  10 &  6.9 &  86 \\ 
    Sm00 &         Smith et al. (2000) &   78 &   75 &  26 &   5 &  5.5 &  42 \\ 
    Hu12 &        Huchra et al. (2012) &   87 &   75 &  -6 &   8 & -0.7 &  70 \\ 
    QS93 & Quintana \& de Souza (1993) &   71 &   58 &  -1 &  16 & -0.1 & 124 \\ 
    Ri87 &              Richter (1987) &   45 &   42 &  67 &  19 &  3.6 & 122 \\ 
    Cr87 &     Cristiani et al. (1987) &   43 &   33 &   1 &  26 &  0.1 & 151 \\ 
    Ko04 &    Koribalski et al. (2004) &   41 &   38 &  25 &   5 &  4.7 &  32 \\ 
    Pi06 &      Pimbblet et al. (2006) &  229 &   35 &  54 &  24 &  2.2 & 143 \\ 
    Og08 &        Ogando et al. (2008) &   39 &   38 &  54 &   5 &  9.9 &  33 \\ 
    La14 &         Lauer et al. (2014) &   44 &   37 &  13 &   7 &  1.8 &  45 \\ 
    Wi91 &       Willmer et al. (1991) &   39 &   36 &  35 &  12 &  2.9 &  72 \\ 
    Me87 &      Metcalfe et al. (1987) &   39 &   37 &  -5 &  14 & -0.4 &  86 \\ 
   The07 &      Theureau et al. (2007) &   31 &   31 &  20 &   9 &  2.2 &  52 \\ 
    Me04 &         Meyer et al. (2004) &   31 &   30 &  13 &  11 &  1.2 &  58 \\ 
   Dal99 &          Dale et al. (1999) &   44 &   29 &  -5 &  15 & -0.3 &  80 \\ 
    Ve90 &     Vettolani et al. (1990) &   43 &   28 & -43 &  20 & -2.1 & 105 \\ 
    MQ81 &  Melnick \& Quintana (1981) &   24 &   12 & -12 &  28 & -0.4 &  96 \\ 
    Jo95 &     Jorgensen et al. (1995) &   23 &   22 &  48 &   6 &  7.6 &  29 \\ 
    Ma14 &       Masters et al. (2014) &   21 &   21 &  23 &  20 &  1.2 &  90 \\ 
    SC98 &      Surace \& Comte (1998) &   35 &   18 &  49 &  37 &  1.3 & 155 \\ 
    Mo15 &       Momcheva et al. 2015) &   27 &   18 &   5 &  14 &  0.3 &  60 \\ 
   Hydra &            This work (CTIO) &  311 &   14 &  98 &  40 &  2.4 & 151 \\ 
    St92 &        Strauss et al (1992) &   16 &   14 &  38 &  15 &  2.5 &  57 \\ 
    PL95 &     Postman \& Lauer (1995) &   15 &   14 & 121 &  15 &  8.0 &  57 \\ 
    dS97 &       deSouza et al. (1997) &   13 &   13 &  12 &  24 &  0.5 &  88 \\ 
    Ku02 &    Kuntschner et al. (2002) &   12 &   12 &  29 &  10 &  2.9 &  34 \\ 
    Ga94 &       Garcia et al. (1994) &   11 &   10 &  20 &  18 &  1.1 &  56 \\ 
\hline
\end{tabular}
\label{zeropoint}
\end{table*}

\subsection{Reported velocity calculation}

The combined velocity catalogue presents one entry per galaxy. The heliocentric velocity assigned to each galaxy is the average of all available values, $v_i$, weighted by their respective errors as reported in the database, $\Delta v_i$. It is given by
\begin{equation}
\overline{v} = \frac{\sum_i {v_i\,\Delta v_i^{-2}}}{\sum_i{\Delta v_i^{-2}}}.
\end{equation}
The error in the combined velocities incorporates both direct measurement errors reported in the original catalogues and the deviation of the individual measurements from the mean, being given by
\begin{equation}
\Delta \overline{v} = \frac{\sqrt{\sum_i{\left[(v_i - \overline{v})^2 + \Delta v_i^2\right]\,\Delta v_i^{-4}}}}{\sum_i{\Delta v_i^{-2}}}.
\end{equation}

Galaxy entries are identified by their SASSSG positions, following the previously discussed cross-identifications. 
These identification codes are consistent with the ones provided in the velocity database, in which each independent velocity measurement is identified using a letter (e.g. G124213.2-275210\_a).
We also list the main SASSSG parameters, particularly the magnitudes, for easy use of this catalogue. A short section of the combined velocity catalogue is shown in Table \ref{velocity_catalog}.

\begin{table*}[htb]
\label{velocity_catalog}

\caption{Shapley Supercluster velocity catalogue.
(1) Galaxy ID;
(2) Right Ascension (J2000); 
(3) Declination (J2000); 
(4) Bj magnitude (COSMOS);
(5) R1 magnitude (COSMOS);
(6) R2 magnitude (COSMOS);
(7) I magnitude (COSMOS);
(8) Object angular area in pixels (COSMOS);
(9) Object major diameter on the focal plane (A$_\mathrm{l}$ [0.01 um], COSMOS);
(10) Object minor diameter on the focal plane (B$_\mathrm{l}$ [0.01 um], COSMOS);
(11) PA parameter in degrees (COSMOS);
(12) COSMOS object type classification;
(13) Heliocentric velocity in $\kms$, corrected by the zero point;
(14) Heliocentric velocity uncertainty in $\kms$, corrected by the zero point error;
(15) Number of spectra retained for an object.
}
\fontsize{7.5}{9}\selectfont
\begin{tabular}{lrrrrrrrrrrrrrr}
\hline
 {\bf(1)}     &    {\bf(2)} &  {\bf(3)}   & {\bf(4)} & {\bf(5)} & {\bf(6)} & {\bf(7)} & {\bf(8)} & {\bf(9)} & {\bf(10)} & {\bf(11)} & {\bf(12)} & {\bf(13)} & {\bf(14)} & {\bf(15)} \\
\hline
 G124213.2-275210 &  12:42:13.255 & -27:52:10.82 &  16.66 &  16.03 &  16.58  & 16.19  &  509 &  15134 &    8601 &   36 &    1    & 2168 &       35  & 2 \\
 G124314.9-310413  & 12:43:14.949 & -31:04:13.29 &  12.28 &  14.86  & 15.76 &  14.90 &   586 &  11619 &    9352  &  58 &    1 &    15673     &   30 &  2 \\
 G124313.7-343419  & 12:43:13.713 & -34:34:19.14  & 16.43 &  16.47 &  16.28  & 16.01 &    651 &  20942  &  7251 &  109   &  1  &   6026 &       42  & 2 \\
 G124351.2-321937  & 12:43:51.224 & -32:19:37.03 &  15.95 &  15.23  & 15.77 &  15.44 &    953 &   17370 &  11232 &    51 &     1    & 9195 &       99 &  2 \\
 G124353.4-322345  & 12:43:53.441 & -32:23:45.91 &  100.00 & 100.00 &  15.56 & 100.00 &  1051 &  16817 &  11807  & 161  &   1  &  16165  &      31 &  1 \\
\hline

\end{tabular}
\end{table*}

\section{Completeness analysis}

We studied the completeness of the galaxy survey by comparing our velocity catalog to the overlapping region of the SuperCOSMOS catalogue (SASSSG) of all galaxies extracted to the same magnitude depth.

The completeness analysis was performed in position/R2-magnitude space, comparing galaxy counts between the merged and SuperCOSMOS catalogs.
The area was divided in $128\times 128$ RA/DEC sections (roughly rectangles), and 50 magnitudes bins ranging from $R2=13$ to $18$, forming a 3-dimensional volume of $128\times128\times50$ blocks.
Then the galaxies in each block were counted.
As the galaxy density was uneven across this 3-dimensional space, we used a variable Gaussian kernel to compute the completeness in each block, smoothing across the grid.
This was done as follows: centered at each block, we grew up an ellipsoid until at least 3 galaxies of the velocity catalog were found inside.
The ellipsoid had a magnitude to angle ratio (deg) of 0.4.
Then, the axes of the ellipsoid was used as the standard deviation of a 3-dimensional Gaussian kernel.
The galaxies were counted up to 3 standard deviations away from the center, weighting them by this kernel.
The same kernel was used to count galaxies in the velocity and SuperCOSMOS catalogs.
The completeness was finally obtained for each block as the ratio between the counts of the velocity and SuperCOSMOS catalogs.
This technique effectively smooths the 3-dimensional region, providing higher resolution in regions with higher galaxy densities.

\begin{figure*}
\begin{centering}
\includegraphics[trim=50 50 100 50, clip, width=1.0\columnwidth]{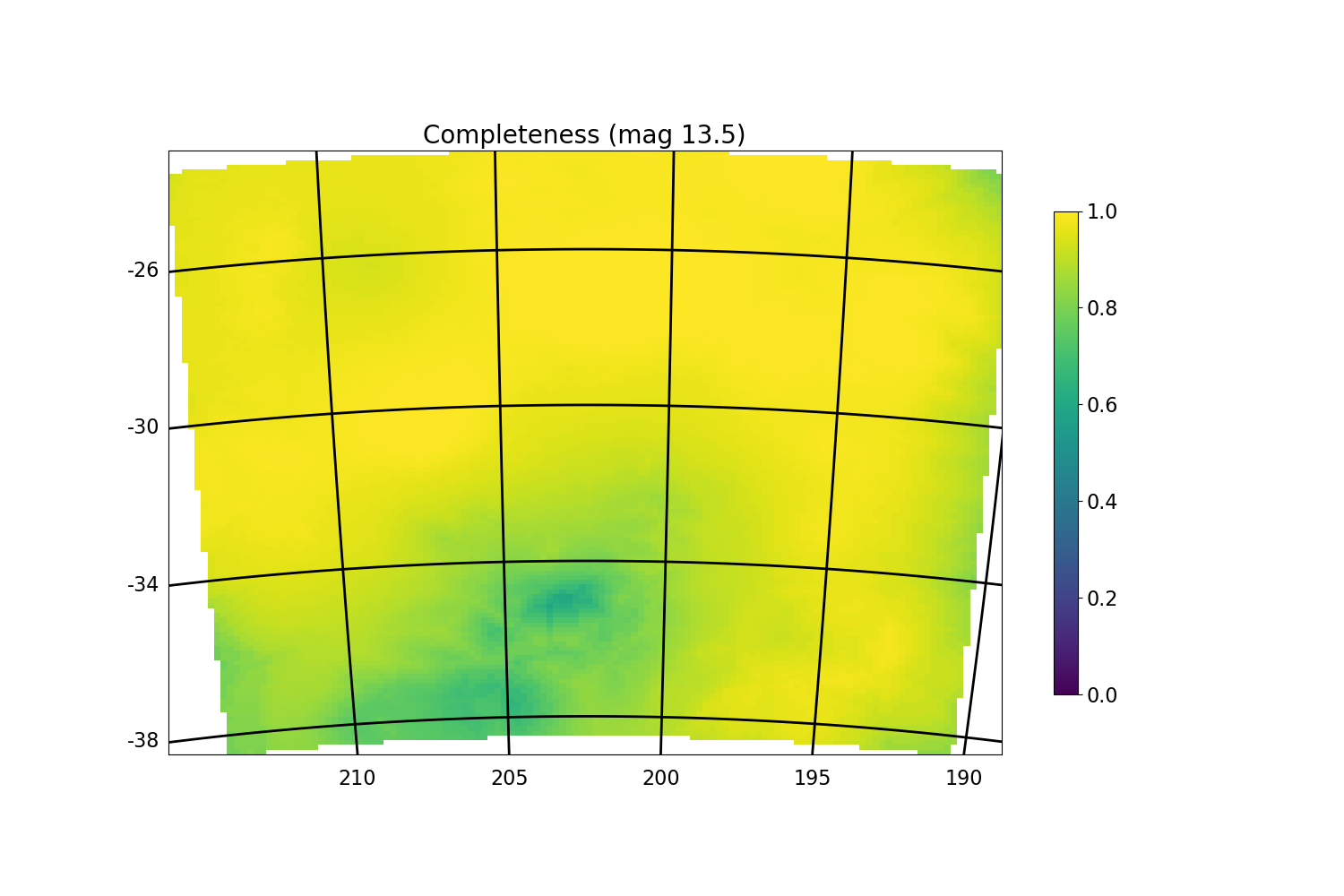}
\includegraphics[trim=50 50 100 50, clip, width=1.0\columnwidth]{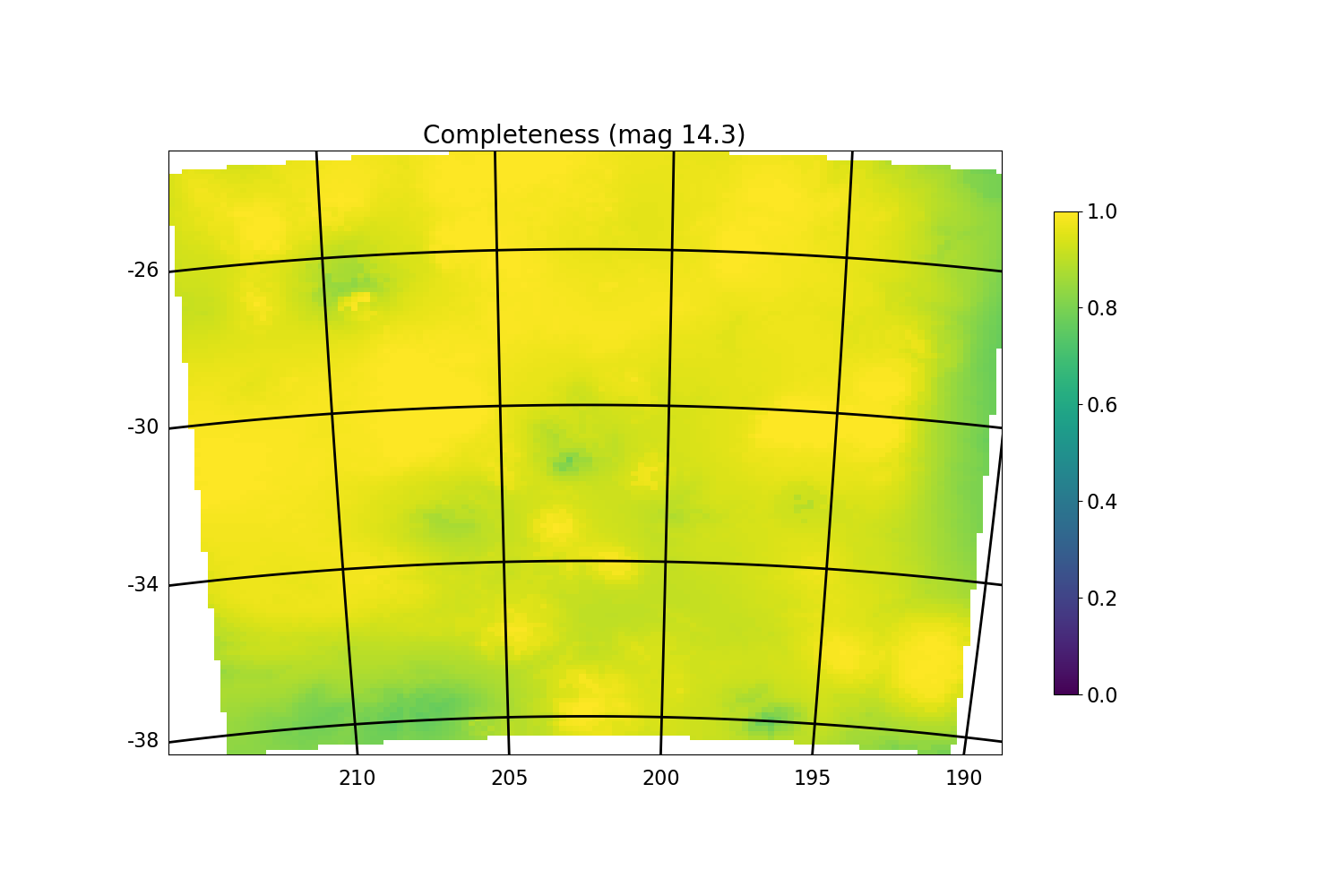}
\includegraphics[trim=50 50 100 50, clip, width=1.0\columnwidth]{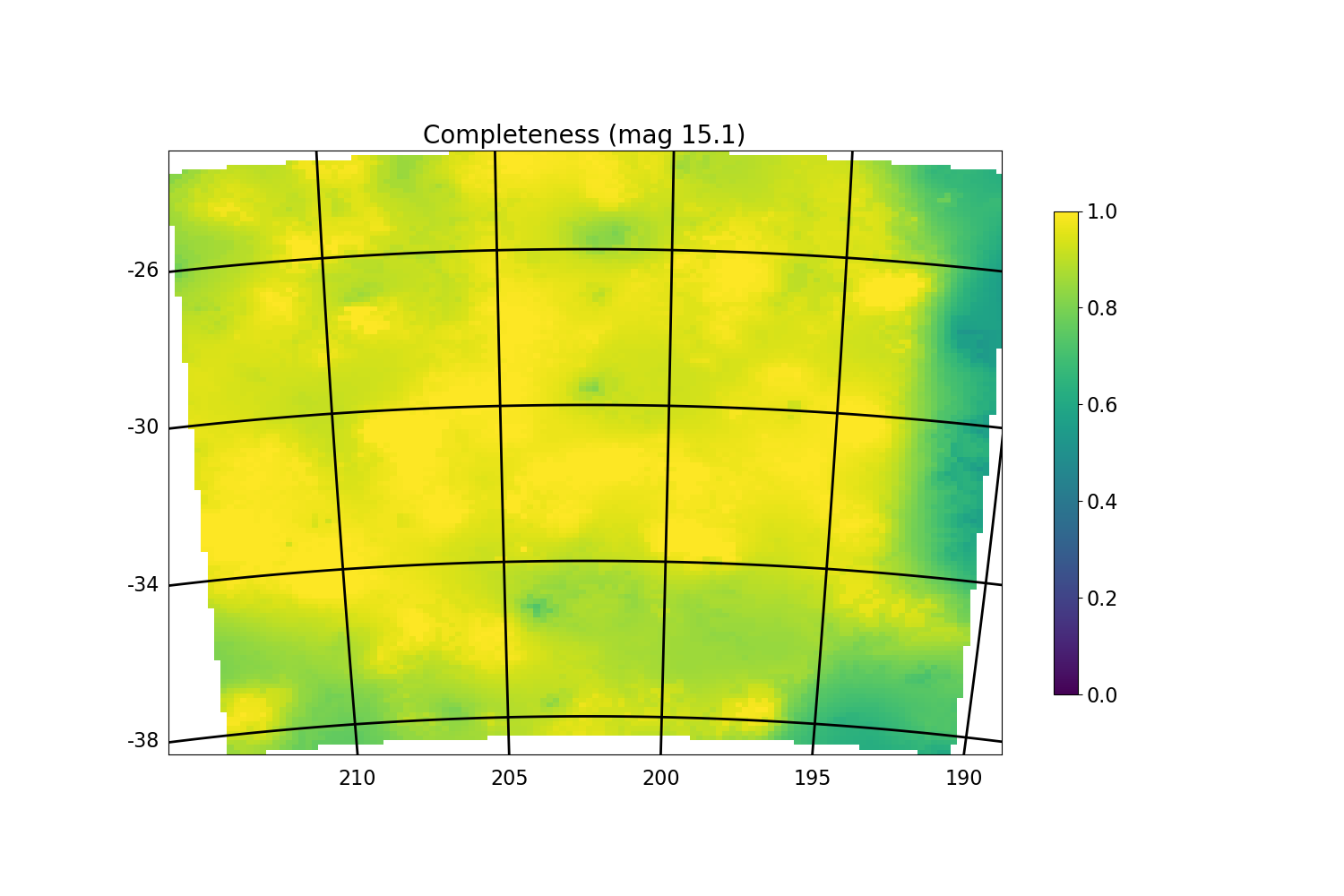}
\includegraphics[trim=50 50 100 50, clip, width=1.0\columnwidth]{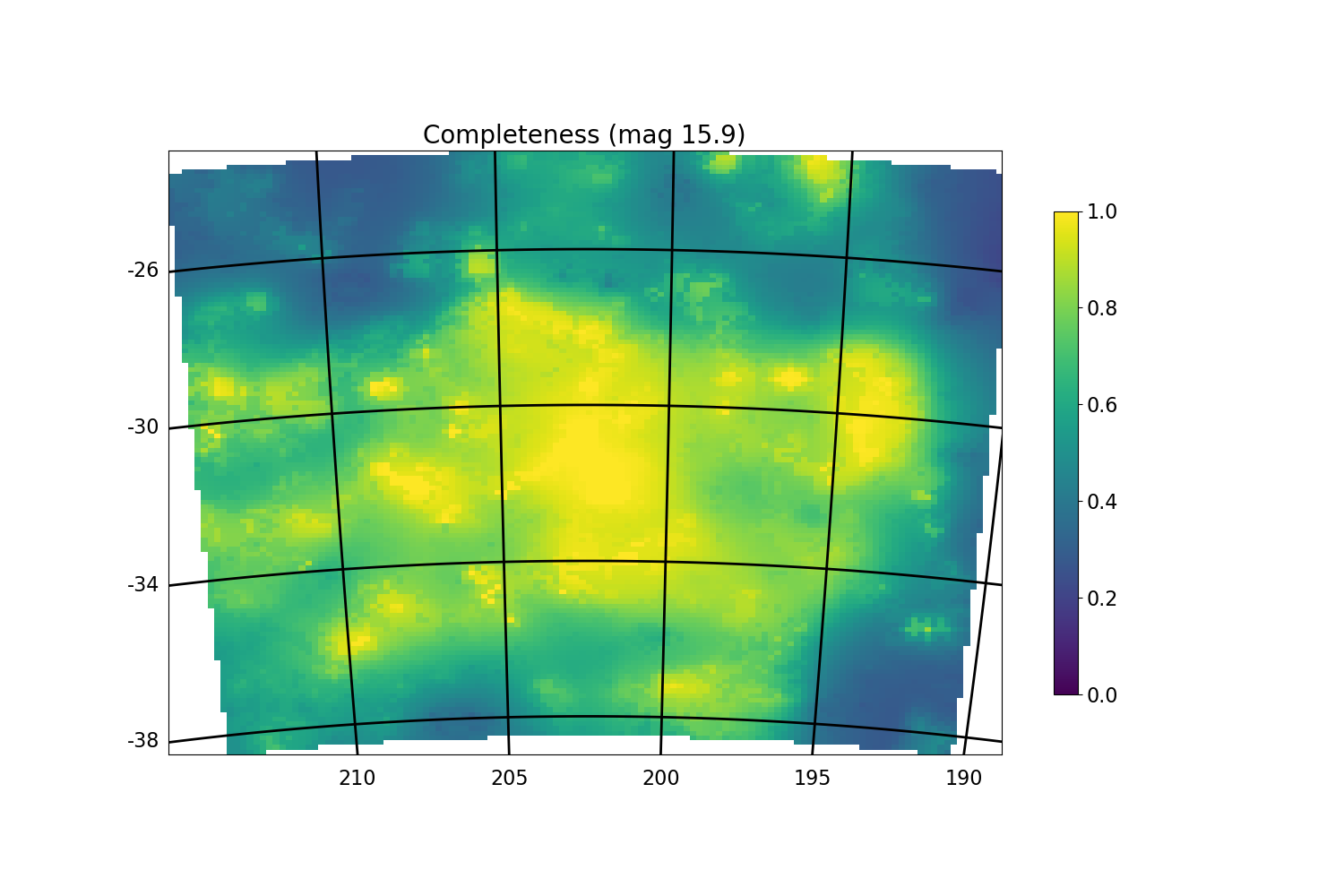}
\includegraphics[trim=50 50 100 50, clip, width=1.0\columnwidth]{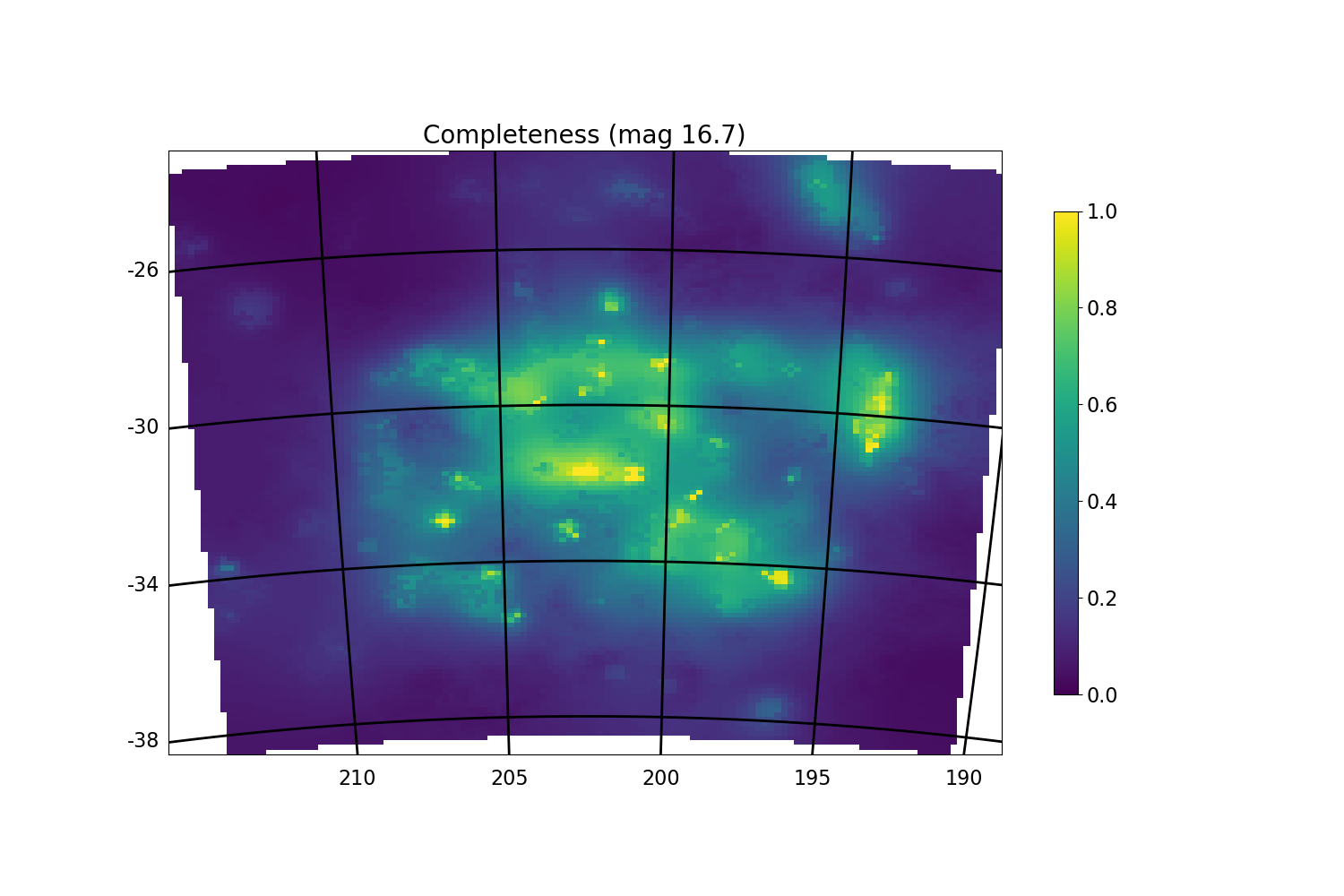}
\includegraphics[trim=50 50 100 50, clip, width=1.0\columnwidth]{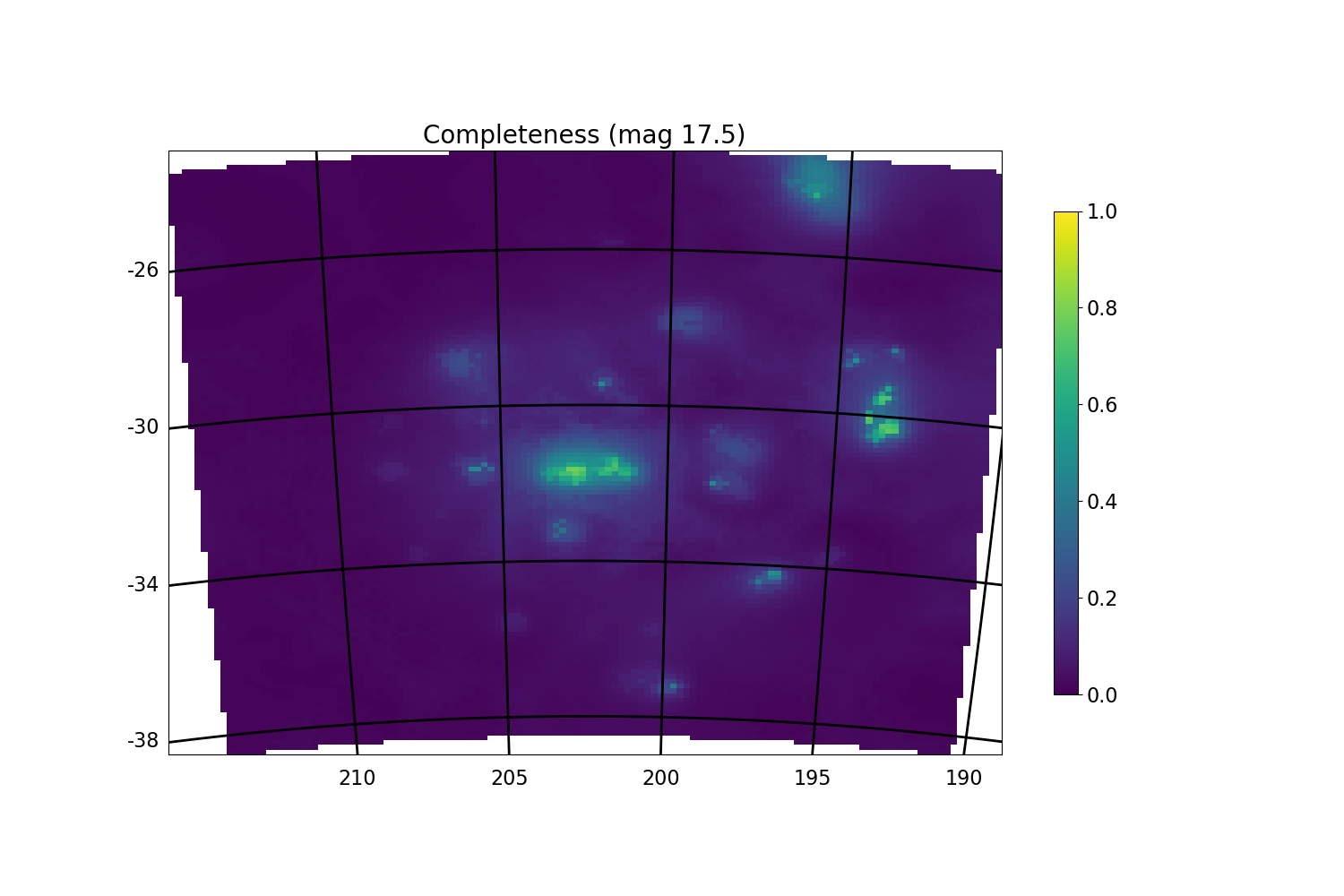}
\par\end{centering}
\caption{Completeness maps for six bins at representative magnitudes. The completeness calculations have been smoothed with a Gaussian filter to ensure that at least 3 galaxies were used (within one sigma) to compute the completeness.}
\label{completeness}
\end{figure*}

\begin{figure}
\begin{centering}
\includegraphics[width=0.9\columnwidth]{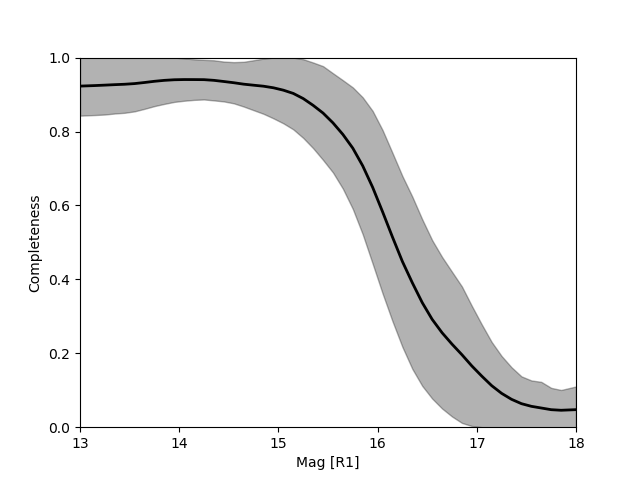}
\par\end{centering}
\caption{Completeness as a function of R2 magnitude for each magnitude bin. The solid line corresponds to the mean and the shaded region shows the standard deviation of the completeness across the studied catalog area.}
\label{completeness_mag}
\end{figure}

Figure \ref{completeness} shows the completeness of the velocity catalog for six magnitude bins.
As expected, the completeness of the catalog is very high at lower magnitudes, and decreases rapidly at magnitudes beyond 15.5, except of the areas near clusters, were the sampling is good up to high magnitudes.
Figure \ref{completeness_mag} provides the completeness as a function of R2 magnitude. 
We stress that this survey did not have a priori completeness figure or limit. 
Several factors produce the incompleteness evident on the figure. 
Among them is the simple addition of in-homogeneous, partial data from the literature of all types of programs, as well as the limited spectra capability of multiple-slits or hole detectors in high galaxy density regions. 
Also, the different capability depth of the several instruments and telescopes used.
Another factor was the lack of an homogeneous photometric catalogue to select spectroscopic targets until the SuperCOSMOS survey was available. 
Anyway, the amount of collected data provides a useful tool to investigate the structure of this supercluster.

\section{Velocity distribution and structure of the Shapley Supercluster}\label{Velocity and topology}

The definition of the structure and topology of the SSC is not an easy task, because of the complexity of the structures in the velocity distribution. 
The presence of many clusters, with their characteristic finger-of-God velocity structures, complicates the study in three dimensions. 
Moreover, remaining irregularities and gaps in the observations could mimic apparent structures. 
Finally, as modern redshift surveys show, dense structures are linked to each other by filaments and walls, forming a fabric that weaves throughout space.

\subsection{Angular distribution}

To improve our understanding of the matter distribution of the supercluster, we generated a three-dimensional density field in redshift-space.
This was done using a 3-dimensional Gaussian kernel in redshift-space, and compensating by the completeness calculations described above. 
Each galaxy was assigned an individual kernel of the form
\begin{equation}
\label{density}
    \kappa(\theta, \phi, v) = \frac{1}{\sqrt{8\pi^3}\,\sigma_a^2\sigma_v}\exp{\left(-\frac{\theta^2+\phi^2}{2\sigma_a^2} - \frac{v^2}{2\sigma_v^2}\right)},
\end{equation}
were $\sigma_a$ is the same angular kernel radius used in the completeness analysis and $\sigma_v$ is the velocity dispersion fixed to 500 $\kms$.
The normalization ensures that each galaxy contributes with a single count when integrating over the full redshift-space volume, provided that $\sigma_a$ and $\sigma_v$ are small compared to the scales of interest.
The density at each point of redshift-space is then computed by summing over the contributions of all galaxies, represented by these (properly shifted) density kernels, weighted by the inverse of the completeness associated to each galaxy.
The units of the resulting density is counts per unit redshift-space volume.
To equalize the pixel solid angles and produce a more physical representation of the density field, we used a flat projection of the sky coordinates onto a tangential plane centered at RA 13:29:36 and DEC -30:59:05.

\begin{figure*}
\begin{centering}
\includegraphics[trim=160 20 50 50, clip, width=2.0\columnwidth]{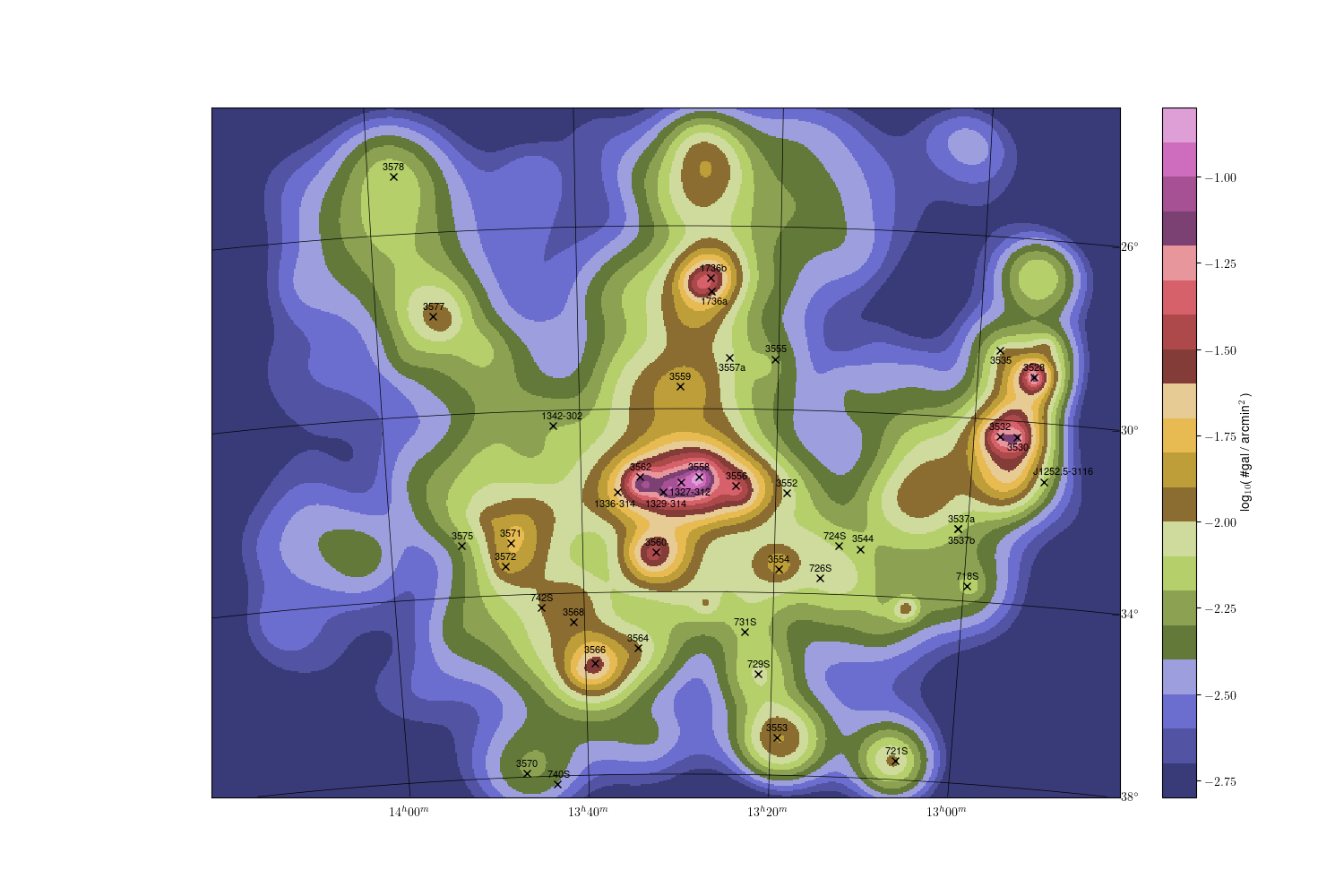}
\par\end{centering}
\caption{Projected density field in the Shapley area with velocities between 9,000 and 18,000$\kms$. Known clusters between 8,500 and 18,500 $\kms$ are included for reference.}
\label{density_field}
\end{figure*}

Figure \ref{density_field} shows a map of the projected density field of the Shapley area.
This was obtained by integrating our redshift-space density field in velocity, such that the units become counts per solid angle in arcminute squared.
The integration ranged between 9,000 and 18,000$\kms$, restricting to SSC relevant structures only.
Note that the redshift-space density field does incorporate matter contributions from structure in front and behind this velocity range, as implied by the kernel in Equation \ref{density}.

\subsection{Velocity distribution and structure of the supercluster}

There are \ngal~galaxies in total, but the importance of the SSC in this region of the sky is demonstrated by the fact that approximately 5600 (52\%) of the galaxies belong to the SSC and its immediate neighborhood, if we consider as such all galaxies with velocities in the range 9,000-18,000$\kms$ (a total depth of 90 $h^{-1}$~Mpc). 
It can be seen that by probing large regions of the SSC away from the richer Abell clusters, we have confirmed significant structures which make links with the main cluster.

\begin{figure}
\begin{centering}
\includegraphics[width=1.0\columnwidth]{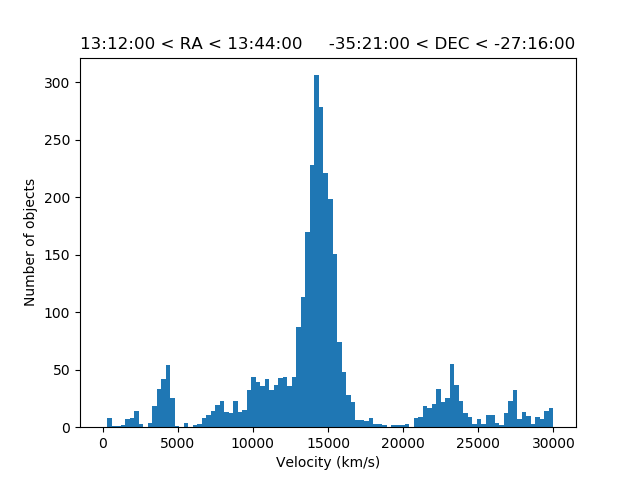}
\par\end{centering}
\caption{Histogram of galaxy velocities in the direction of the Shapley Supercluster with all velocities available in the range $0\leq v \leq$ 30,000$\kms$, with a step size of 300$\kms$.}
\label{vel_hist}
\end{figure}

\begin{figure*}
\begin{centering}
\includegraphics[width=\columnwidth,valign=b]{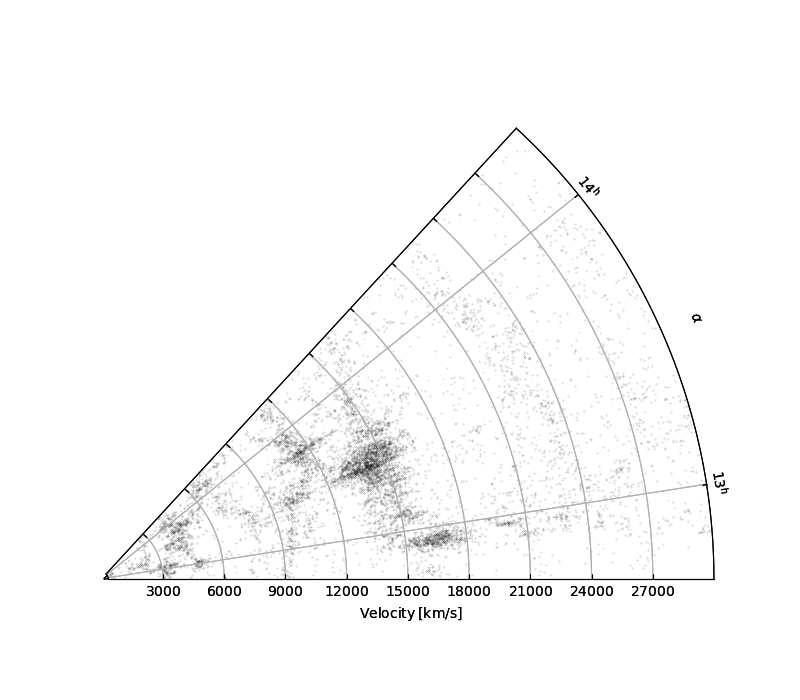}
\includegraphics[width=\columnwidth,valign=b]{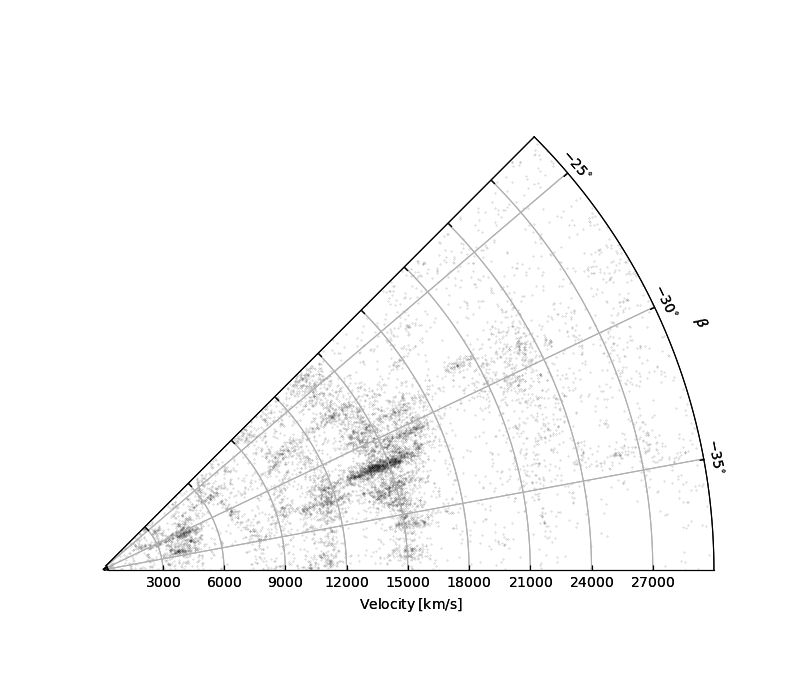}
\par\end{centering}
\caption{Two projections in R.A. (left) and Dec (right) of the distribution of galaxies with measured redshifts in
the region of the Shapley Supercluster until 30,000$\kms$. The angle in Right Ascension is expanded by a factor 2 and in declination by a factor~3 relative to its true size for clarity.}
\label{vel_proj}
\end{figure*}

\begin{figure*}
\begin{centering}
\includegraphics[width=\columnwidth,valign=b]{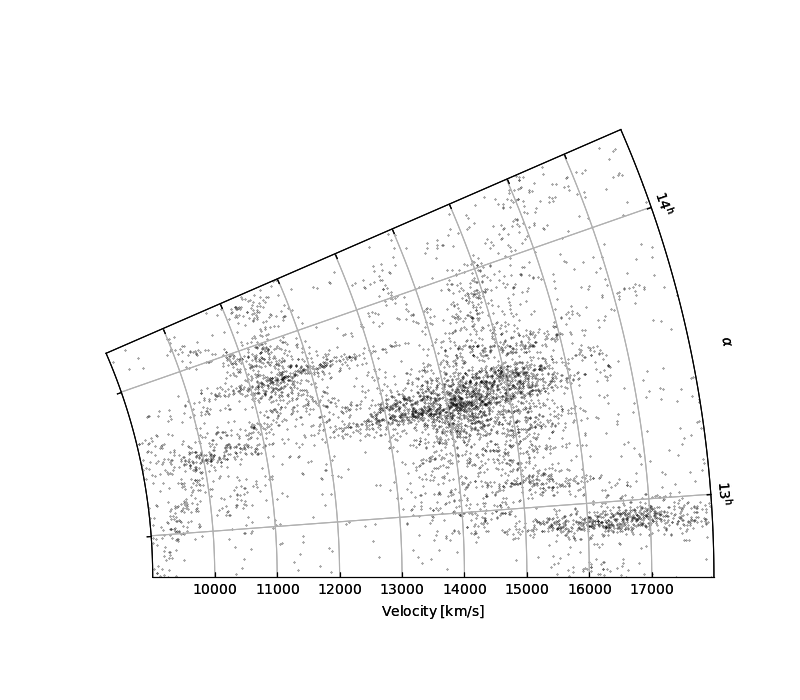}
\includegraphics[width=\columnwidth,valign=b]{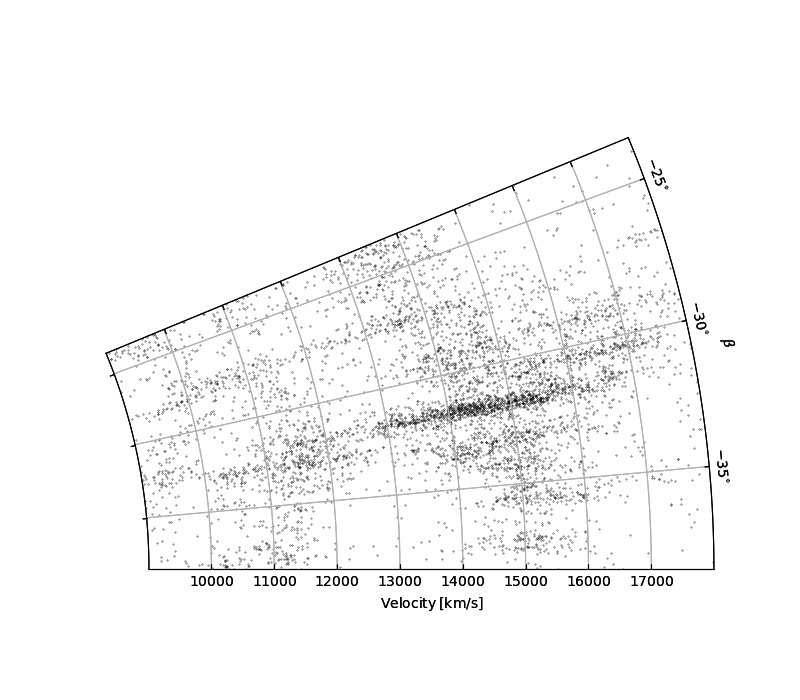}
\par\end{centering}
\caption{Two projections in R.A. (left) and Dec (right) of the distribution of galaxies with measured redshifts in
the region of the Shapley Supercluster with velocities between 9,000 and 18,000$\kms$. The angle in declination is expanded by a factor~1.5 relative to its true size for clarity.}
\label{vel_proj_shapley}
\end{figure*}

From the density plot and the velocity distribution, we can infer some features of the 3-d structure of the supercluster, taking due consideration of the abundant presence of fingers of god in this dense volume. 
Figure \ref{vel_hist} shows the histogram of the velocities of galaxies in the direction of the SSC with all available velocities in the range $0 \leq v \leq$ 30,000$\kms$, with a step size of 300$\kms$. The histogram shows several maxima, as discussed by Quintana \etal (2000) and Proust \etal (2006).
Figure \ref{vel_proj} shows the combined resulting distribution of galaxies towards the SSC as wedge diagrams in Right Ascension (left) and Declination (right) of the whole velocity catalogue until 30,000$\kms$. 
Figure \ref{vel_proj_shapley} is a similar plot that provides a closer look are the SSC region, limited between 9,000 km/s and 18,000 km/s, revealing the rich interconnected structure within the central parts and to their neighbouring closer concentrations.

The prominent foreground wall of galaxies (Hydra-Centaurus region) is defined centered close to 4,200$\kms$. A second foreground structure at approximately 7,200$\kms$ is composed by galaxies forming a tail between the Hydra-Centaurus and Shapley regions, as can be seen in the Wedge diagrams of Figure  \ref{vel_proj}. 
The main body of the SSC is represented by the highest peak of 4,600~galaxies, which is centered at approximately 15,000$\kms$. 
A few low relative peaks centered between 10,000$\kms$ and 13,000$\kms$, forming a sort of plateau in the histogram, shows the nearer concentration which is located to the East, centered on A3571 and connected to the main SSC body, as shown in Figure \ref{vel_proj_shapley}.

The larger velocity catalogue used here confirms the general structure and the main features of the SSC already discussed in Quintana \etal (2000) and Proust \etal (2006). 
For completeness we briefly summarize them here. 
The central region is roughly spherical in shape centered on the cluster A3558, and has at its core the highest-density, elongated volume containing the Abell clusters A3562, A3558 and A3556 with almost identical recession velocities around 14400$\kms$ and the groups SC1329-314 and SC1327-312, whose more discrepant velocities (by several hundred kilometers per second) could be attributed to the in-fall component along the line of sight.
Towards the south of the elongated feature, the central region contains also the cluster A3560. 
As described in Reisenegger \etal (2000), the whole of this central region and all of its immediate surroundings are within the volume that is currently undergoing gravitational collapse. As shown in Figure \ref{vel_proj}, we note the presence of the prominent foreground wall of galaxies of the Hydra-Centaurus region. Moreover the ``Front Eastern Wall'' (Quintana \etal 2000), composed with a bridge of galaxies, groups and clusters, extends to the east and in front of the supercluster, the densest part being at $\simeq$ 10,000-11,000$\kms$, located to the east. It contains the clusters A3571, A3572, A3575 and the group SC1336-314.
The A3570 cluster is located at the southern tip of the observed part of the wall and A3578 at its northern one. This wall establishes a further link between the Hydra-Centaurus region and the SSC, while a second one extends towards the west at $\overline{v}= 8300\kms$. 
Clumps of objects clearly link the two main structures. 
However, care must be taken in the interpretation of the wedge plots because of the finger-of-God effect evident in the main SSC concentrations (made especially prominent by the higher completeness in these cluster regions) and because of an analogous effect with opposite sign due to the inflow on larger scales, which makes the overdensities appear more overdense in redshift space than they are in the real space.

West and slightly North of the SSC core, another rich concentration of galaxies is connected to the central SSC regions. The clusters A3528, A3530 and A3532 form a concentration of galaxies and clusters at about R.A.\,=\,12h50m and $v=$ 16,000-17,000$\kms$ connected to the main body of the SSC by a broad bridge of galaxies. 
It can also be seen from the wedge diagram in declination (Fig. \ref{vel_proj_shapley}) that the southern part of the SSC consists of several clouds of galaxies where the known Abell clusters represent the peaks of maximum density. 
In this diagram, the sheet at $\overline{v}=$ 15,000$\kms$ is present right across the observed region from $-23^\circ$ to $-38^\circ$, which corresponds to the long complex of galaxies as shown on the Fig. 9 of the 6dF~survey (Heath Jones \etal 2009). 
Note that the central region of the SSC with 11 clusters (A3552, A3554, A3556, A3558, A3559, A3560, A3562, AS0724, AS0726, SC1327-312 and SC1329-313 has been also analysed by Haines \etal (2018) leading to similar conclusions. They give evidence that A3560 has two distinct sub-structures within ${\it r_{200}}$ one to the north and one to the west and its connecting structure is NW towards A3558.


\section{Conclusions}
We have presented a catalogue containing \nvel~velocities for \ngal~galaxies in the area of the SSC by combining existing measurements from the literature with \newvel~new velocities for \newgal~new galaxies observed by us at Las Campanas and CTIO observatories.
The objects were cross-matched with galaxies from the SuperCOSMOS Sky Surveys to incorporate proper photometry and astrometry to the velocity measurements.
This compilation of catalogs from various authors resolves several issues that we encountered, like coordinate misprints, zero-point velocity corrections, and cross-matching identification errors, resulting in a clean catalog for scientific use.
We provide the catalog together with a completeness estimation as function of magnitude and position on the sky across most of the catalog area.
All these data products are now available for the scientific community in digital format, and can be accessed here: \url{http://www.astro.puc.cl/Shapley-catalogs}.

By adding velocity measurements covering areas between known clusters, we improved our understanding of filaments connecting the central clusters, revealing a complicated structure of gravitationally interacting elements.
These elements are yet to be proven gravitationally bound, an analysis that we are leaving for a future publication.
The extended catalog also reveals the possible existence of several new groups and poor cluster candidates, also allowing to test the reality of some proposed structures, which will also be explored in a future study.

\appendix

\section{HI velocity measurements and possible wrong or confused identifications.}

We found a number of velocity and/or positional discrepancies between optical and HI measurements. For these cases there are 3 possible explanations: 
a) the optical velocity is wrong due a number of factors, either misprints, 
wrong identification of the galaxy at the telescope, or just a wrong velocity determination (a blunder). 
b) the HI velocity is the result of a confusion or reflection from other sources in the beam. 
c) Since the HI positions are uncertain typically between 1'-2', and up to 6' (see HIPASS optical catalogue, HOPCAT, Doyle et al 2005), there could be a wrong identification of the HI source.

When there are two or more consistent and independent optical velocities, the possibilities are reduced to b or c. Therefore, the correct identification is given by the associated to the optical velocity based on the galaxy image. One factor affecting the reliability of the identifications is caused by the allocation of the HI velocity to a 
bright galaxy near the HI position, which is identified with the source. Different catalogues have adopted this practice (FLASH, NED, Simbad, possibly ZCAT, others?), 
without keeping the original HI pointing coordinates (sometimes, these are unpublished). Furthermore, due to the small velocity errors of the HI measurements, these velocities are selected by previous catalogs as the most precise $z$ values.

In this Appendix we discuss eight cases of discrepant HI vs. optical velocities, with comments concerning possible identification alternatives. Two galaxies, ESO 507-036 and IC4254, have optical redshifts that place them in the Shapley supercluster velocity range, further away than normally listed.

ESO 381-020, a low-surface-brightness (disaggregated) Irr galaxy. Data is \#22 in Database (and comments). NED and other catalogues give a low $z$ derived from HI velocities.
Problem: Dr91 gives 3,399$\kms$, while Ka03 (source ZCAT from HI data) and Ko04 (HIPPASS) give ~585$\kms$, discrepant velocities from Dr91. All 3 have positions consistent with ESO381-020. The velocity quoted by Ka03 may not be an independent measurement. 
The Cosmos galaxy image has morphology consistent with $v\approx$600$\kms$. Its large angular extension means the 3 positions are spread over the central very irregular area. There are no other nearby Sp or Irr galaxies within 15'. Either the Dr91 measurement is faulty, or the Dr91 velocity corresponds to a background galaxy, hard to distinguish among irregularities of the main lsb Irr galaxy image.

ESO 507-036, an Sp galaxy. Data is \#8463 in Database (and comments). The 6dFGS survey gives a velocity of 13,803$\kms$ and a well centered position on the Sp nucleus. Huchtmeier \etal (2005, Hu05) gives an HI value of 3,364$\kms$ with a position within the optical edge-on Sp image. Another 6dFGS Sp at 13,551$\kms$ is located $\approx$5.5'to the north. 
Both publications give the coordinates of the ESO galaxy, with highly discrepant velocities, but 6dFGS is an exact direct coordinate measurement.
The Cosmos optical image of the edge-on spiral suggests a warp of the main disk with a hotspot, or maybe it is a possible background, also edge-on, spiral aligned with the main disk. However, the most likely option is that the Hu05 value is a velocity of a different Sp, as there are several smaller spirals within 5', and there is another edge-on Sp with 3,000$\kms$ at $\approx$20' to the SE.

ESO 443-059, a bright spiral. Data is \#1472 in Database (and comments). Three optical velocities (6dFGS, Dr91, Da87) and one HI (Maia \etal (1996, Ma96 in NED) are consistent at $\approx$3,400$\kms$ 
and consistent with the ESO443-059 position.
Problem: the FLASH survey Ka03 quotes a NED HI source that gives 2,293$\kms$. The discrepant FLASH velocity indicates a wrong id for the HI source. 
Most likely the source could be identified with Database \#1492 Sp at $\approx$3' to the NE, which has 3 velocity measurements around ~2,310$\kms$. Alternatively, but less likely, 
the Ka03 velocity could be identified with \#1507 with v$=2266$ (velocity from Ba97) located $\approx$9' away to the NE. Both are likely members of a group around nearby NGC 4965 at $\approx$2,300$\kms$.

ESO 443-061, a spiral. Data is \#1504 in Database (and comments). There are two independent and consistent optical velocities (6dFGS and LC97) at $\approx$9,500$\kms$. The Ko04 HIPASS brightest-1000 survey measures a 
velocity of 2,282$\kms$ from a position 5'15" away from this galaxy. NED and others 
assign this Ko04 HI velocity to ESO 443-061, which is a wrong optical identification. The HIPASS velocity is consistent with some NGC 4965 group members. Alternatively, 
it could be a rare HI cloud, or another galaxy. In fact, the most likely identification for this HIPASS source is with \#1507 from our Database. At 2'10" this is closer to the HIPASS position than to ESO 443-061.

IC 4254 a bright peculiar galaxy. Data is \#3681 in Database (and comments). Two optical, independent and consistent velocities (Dr88, St96) give $\approx$15,350$\kms$ and a make it a likely member of the A1736 cluster (old velocity NED value used by Ka03). 
However, HI observations at Nancay (The98) give a velocity of 3,503$\kms$.
Problem: The98, NED and many catalogues give IC4254 the low z, but the optical distance is $\approx$5 times higher, in a rich cluster within the SSC.  
Alternative identifications for the Nancay The98 observation: few nearby galaxies are without velocity measurements (the field is that of the A1736 cluster). Possibilities are an Sp 2'20" to N (at 13 27 44.11 -27 11 03.3; R2$=16.47$, though 
it looks distant), and more likely, an lsb Irr gal 4'10" to NW (at 13 27 30.03 -27 10 27.5; R2$=16.8$)

ESO 444-082. Data is \#5238 in Database (and comments). With 3 consistent positions, a very low HI Nancay velocity of $\approx$526$\kms$ is given by The98, while 2 independent consistent (QC03, 6dFGS) optical measurements give a value $\approx$11,300$\kms$. 
NED and others chose the low z from The98, which in view of the optical data looks like a wrong identification or bad data. It is unclear if the Nancay position is just that of the optical image, with a wrong identification. 
Alternatives: No obvious candidate within 10'. No obvious explanation. Perhaps confused HI data.

NGC 5298, a bright Sp. Data is \#6234 in Database (and comments).  Nancay The98 HI velocity at 4,083$\kms$ and identification (Ka03). 
Four optical velocities (LC97, Dr91, Da87 and 6dFGS) and one HI velocity (Mathewson \etal 1992, Ma92) are consistent at $\approx$4,400$\kms$, and the position is in A3574 cluster field, at similar z (4,400$\kms$).
However, NGC 5298 velocity observed by Nancay (The98) has a discrepancy of 350$\kms$, quite significant for an HI measurement. It seems the galaxy identification could be misplaced, as
there are a few surrounding spirals with closer velocities to The98 4,083$\kms$: closest is 1' to N.

Galaxy L606P Sp. HI Nancay (Monnier-Ragaine \etal 2003, Mon03) gives a 541$\kms$ velocity and identification. Data is \#6839 in Database (and comments). There is a 6dFGS optical velocity measured of 3,778$\kms$ of this fairly small Sp. Several nearby galaxies have similar z, which is of cluster S0753. 
The Nancay 541$\kms$ low velocity (it should be a very near galaxy), is discrepant value to 6dFGS. Several surrounding brighter galaxies at 3,800-4,400$\kms$ (in S0753). No likely candidate within 15'. It could be a confused source or an HI cloud.

\begin{acknowledgements}
HQ is grateful to Las Campanas and Cerro Totolo Interamerin Observatories for generous allocations of telescope time. DP thanks the Instituto de Astrof\'isica of  Pontificia Universidad Cat\'olica de Chile and ESO in the context of the {\it Visiting Scientists program} for their hospitality in Santiago (Chile). RD and AR acknowledge support from CONICYT project Basal AFB-170002 (CATA). This research has made use of data obtained from the SuperCOSMOS Science Archive, prepared and hosted by the Wide Field Astronomy Unit, Institute for Astronomy, University of Edinburgh, which is funded by the UK Science and Technology Facilities Council.
\end{acknowledgements}


\end{document}